\documentclass[10pt,a4paper]{article}

\usepackage{isorot}

\usepackage{graphicx,subfigure}

\usepackage[square,numbers]{natbib}

\usepackage{colortbl}

\usepackage{booktabs}
\usepackage{multirow}

\usepackage{pstricks,pst-plot}

\usepackage{amsmath}
\usepackage{amsthm}
\usepackage{amsfonts}
\usepackage{amssymb}
\usepackage{ipa}
\usepackage{latexsym}
\usepackage{stackrel}
\usepackage[T1]{fontenc}
\usepackage{enumerate}
\usepackage{fourier}
\usepackage{skak}

\usepackage{lastpage}

\newtheorem{teo*}{Theorem}

\title{Mortality Models based on the Transform $\log(-\log x)$}

\author{Meitner Cadena\thanks{Correspondence to: Meitner Cadena. UPMC Paris 6 \& CREAR, ESSEC Business School;\, E-mail: meitner.cadena@etu.upmc.fr, b00454799@essec.edu, or meitner.cadena@gmail.com}} 

\date{}

\begin{document}

\maketitle

\begin{abstract}
A new stochastic method for describing mortality is proposed and explored.
It is based on differences of observed times series of the transform $\log(-\log x)$ of survival probabilities which seem to follow simple patterns over the years.
These common structures are gathered by a representation based on age-constants and time-stochastic processes.
From the projection of the time-processes the mortality forecasting is straigthforward.
Comparisons of the new model with the well-known Lee-Carter and Cairns-Blake-Dowd models employing sex-based mortality data of some countries are provided.
Some in-sample and out-of-sample goodness-of-fit criteria show that 
in some situations the new model performs better than the ones mentioned above.
Assessments of the performance of these models using rates of mortality improvement are discussed.

\vspace{2ex}


\vspace{2ex}

\end{abstract}

\section{Introduction}

Over the last decades, general populations have known sharp improvements in their mortality.
These movements for all ages have been largely attributed to the technical development in public health and improvements in socio-economic conditions (see e.g. \cite{AntolinBlommestein2007}, \cite{DeatonPaxson2004} and \cite{SOA2011}).
This evolutionary phenomenon has impacted domains that are critical for the functioning of societies, as for instance planning for social
security and health care systems and for the funding of retirement income systems, and employment and education organization for increasing older populations.
Hence adequate descriptions of the future mortality are needed.

Growing efforts to provide appropriate mortality forecasts have been known after mainly the seminal paper of Lee and Carter \cite{LeeCarter1992} in 1992.
They introduced a stochastic approach to forecast mortality of a population using age and calendar year.
Their innovative proposal is still attractive and used extensively because of reasonable fits for most Western countries and simplicity and ease of use in practice,
for instance the Lee-Carter methodology is used as a benchmark methodology by the US Bureau of the Census.
Moreover, the Lee and Carter model has motivated to further authors the development of extensions and variants, for instance Brouhns et al. \cite{BrouhnsDenuitVermunt2002}, Li and Lee \cite{LiLee2005}, Renshaw and Haberman \cite{RenshawHaberman2006}, and Hyndman and Ullah\cite{HyndmanUllah2007}.
Other models have been also developed to deal with the challenges that future mortality presents.
Surveys of techniques dealing with mortality projections are presented in e.g. \cite{Pitacco2004}, \cite{PitaccoDenuitHabermanOlivieri}, and \cite{BoothTickle2008}.

Literature shows that some of those models have been based on transforms of survival functions.
This type of models consists in transforms that are applied to survival functions in such a way that their outcomes have much simpler dynamics than those of survival functions.
In this way, these models may describe an important part of the mortality data variability over fitting periods as well as over forecasting periods.
Among well-known models of this type are the relational method based on the logit transform introduced by Brass in 1974 \cite{Brass1974}, and, more recently, inspired by the Wang transform, the model proposed by de Jong and Marshall in 2007 \cite{deJongMarshall2007} which describes shifts of z-scores. 
In these two models the transforms of survival functions are represented by principally time parameters.

Let us see closer these models based on transforms of survival functions, which are related to our new model.
Brass proposed the model (see \cite{Brass1974} and e.g. \cite{PitaccoDenuitHabermanOlivieri})
\begin{equation}\label{eq:20150117:001}
\textrm{logit}(S(x,t))=\alpha_t+\beta_t\ \textrm{logit}(S(x,\tau))\textrm{,\quad$t>\tau$,}
\end{equation}
where $S(x,t)$ is a survival function at age $x$ and in year $t$, logit$(x)=\log\big(x\big/(1-x)\big)$, $0\leq x\leq1$, is the logit transform of $x$, $\log(x)$ represents the natural logarithm of $x$, $\alpha$ and $\beta$ are constants, and $\tau$ is a given year.
On the other side, de Jong and Marshall proposed a model based on z-scores given by (see \cite{deJongMarshall2007})
$$
z_{x,t}=\Phi^{-1}(S(x,t))=R'_x\beta+\alpha_t+\epsilon_{x,t}\textrm{,\quad$d\alpha_t=\lambda d+\sigma db_t$,}
$$
where the first part contains the standard normal distribution $\Phi$ and is a linear regression model with a known regressor vector $R_x$, $\beta$ an unknown regression parameter vector, $\alpha_t$ a value to be defined in
the second part, and $\epsilon_{x,t}$ zero mean measurement errors, and the second part models $\alpha_t$ as a Wiener process with drift $\lambda$ and variance $\sigma^2$ ($\sigma > 0$). 
The Wang transform $\Phi(\Phi^{-1}(x)+\alpha)$ is identified when these authors rewrite the continuous model in its discrete version and introduce their observation that the shifts of the z-scores (obtained from the application of the inverse of the standard normal distribution to survival functions) between successive years seem to be constant over time, which gives
$$
S(i,n+k)=\Phi(z_{i,n}+{\lambda}k)=\Phi(\Phi^{-1}({S}(i,n))+{\lambda}k)\textrm{.}
$$
The following rewriting of the last expression is of interest for us:
\begin{equation}\label{eq:20150117:002}
\Phi^{-1}({S}(i,n+k))=\Phi^{-1}({S}(i,n))+{\lambda}k\textrm{.}
\end{equation}
Note that \eqref{eq:20150117:001} and \eqref{eq:20150117:002} are relationships among transforms of survival functions.
Also, the dynamics of these models in terms of time are reduced to the dynamics of the parameters $\alpha_t$ and $\beta_t$, and ${\lambda}k$ respectively.

We consider in this article the application of the transform $L(x)=\log(-\log x)$, $0<x<1$, to survival functions for representing and forecasting mortality.
To this aim, the differences of these applications of $L$ between any year and another given year are modeled by using time- and age-parameters.
The resulting relationship is like \eqref{eq:20150117:001} or \eqref{eq:20150117:002} and thus the mortality forecasting is reduced to the forecasting of the time-parameters.
Hence mortality forecasting is straightforward.

This paper is organized as follows.
The application of $L$ to survival functions is examined in Section \ref{examination}.
Section \ref{model} presents the formulation of our new stochastic mortality model based on $L$.
Section \ref{illustrations} presents numerical illustrations of applications of the new model to sex-based mortality data of some countries.
Comparisons of these models with the well-known Lee-Carter (LC) and Cairns-Blake-Dowd (CBD) models using in-sample and out-of-sample goodness-of-fit criteria of mortality projections are provided.
Additionally, assessments of the performance of these models, using rates of mortality improvement, are discussed.
The last section presents conclusions and next steps of future research with the new proposed model.

\section{Examination of the application of $L$ to survival functions}
\label{examination}

In this paper survival functions are intensively used.
We first present some ways to obtain these functions since they are not available in life tables.
To this aim, we will use the probability that an individual aged $x$ in year $t$ does not reach $x+1$, $q_{x,t}$, and the central death rate, $m_{x,t}$.
For simplicity, we suppress the explicit dependence on $t$ in all notations when no possible confusion.

On the one side, for a given age $x_0$, the survival function $S(x)$ associated to $x_0$ is based on the probabilities that someone aged exactly $x_0$ will survive for $x-x_0$ more years,
then die within the following 1 year.
These probabilities are classically written as $_{x-x_0|1}q_{x_0}$ and for ease of notation we write them as $r_x$.
In the actuarial literature, the curve of $r_x$, $x$ = $x_0$, $x_0+1$, \ldots, is called the \emph{curve of deaths}.

$r_x$ and $q_x$ are related by:
$$
r_x=
\left\{
\begin{array}{ll}
q_x & x=x_0 \\
q_{x}\times\prod_{i=x_0}^{x-1}\left(1-q_{i}\right) & x>x_0\textrm{,}
\end{array}
\right.
$$
and $S(x)$ corresponds to
$$
S(x)=\sum_{k\geq x}r_k\textrm{,\quad $x\geq x_0$.}
$$
It is not hard to see that $\displaystyle \sum_{k\geq x}r_k=1$ and $\displaystyle S(x)=\prod_{i=x_0}^x(1-q_i)$, $x\geq x_0$.
Conversely, from $r_k$ or $S(x)$ one can compute $q_x$ using
$$
q_x=
\left\{
\begin{array}{ll}
r_x & x=x_0 \\
\frac{r_x}{\prod_{i=x_0}^{x-1}\left(1-q_{i}\right)} & x>x_0\textrm{,}
\end{array}
\right.
$$
or $q_x=r_x\big/S(x-1)=1-S(x)\big/S(x-1)$, $x>x_0$, recalling that $r_{x_0}=q_{x_0}$.

On the other side, throughout this paper, we assume that the force of mortality, defined by $\mu(x)=-S'(x)\big/S(x)$, is constant within bands of age and time, but
allowed to vary from one band to the next, i.e.
\begin{equation}\label{eq:20150120:001}
\mu_t(x)=\mu_{t+\epsilon}(x+\delta)\textrm{\quad for all\quad $0\leq\epsilon,\delta<1$.}
\end{equation}
This implies that $\mu_t(x)=m_{x,t}$ and $e^{-\mu_t(x)}=1-q_{x,t}$.
Combining these relationships gives $e^{-m_{x,t}}=1-q_{x,t}$, which allows the computation of survival probabilities from $m_{x,t}$ using the previous relationships between $r_x$, $q_x$, and $S(x)$.

Assumption \eqref{eq:20150120:001} is often used in mortality data analyses, see e.g. \cite{BrouhnsDenuitVermunt2002} and \cite{CairnsBlakeDowdCoughlanEpsteinOngBalevich2009}.

\begin{figure}[!ht]
\centering
\subfiguretopcapfalse
\subfigure[Curve of deaths ($r_{x,t}$)]{
\includegraphics[scale=0.70]{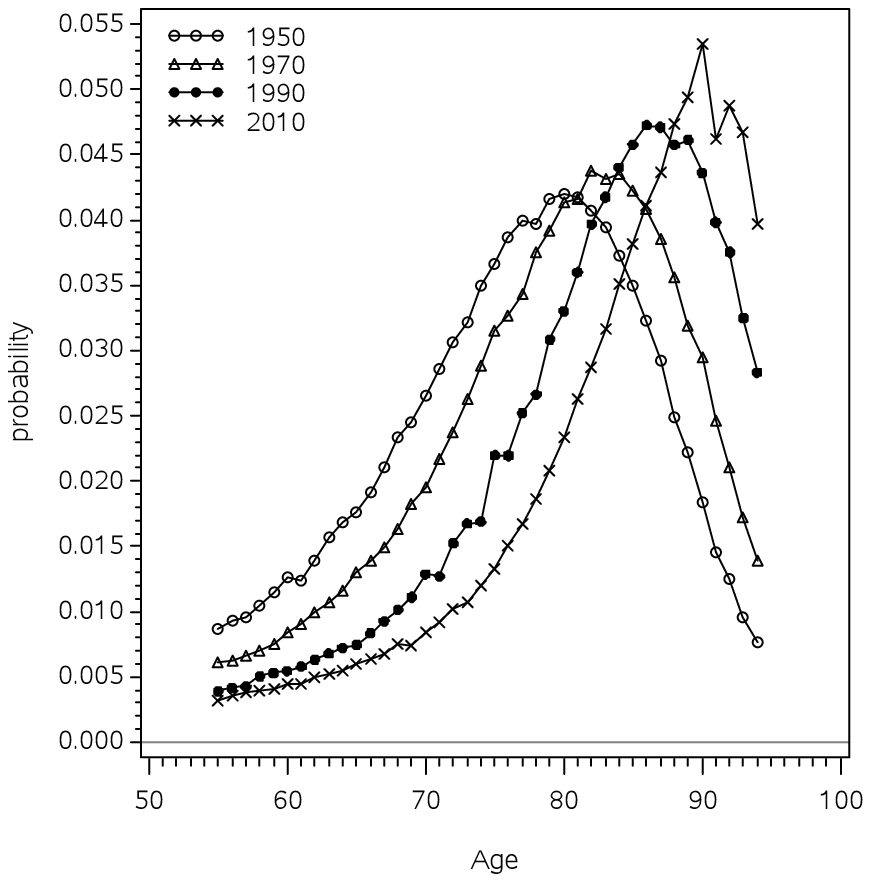}
}
\hspace{1mm}
\subfigure[Survival function ($S_t(x)$)]{
\includegraphics[scale=0.70]{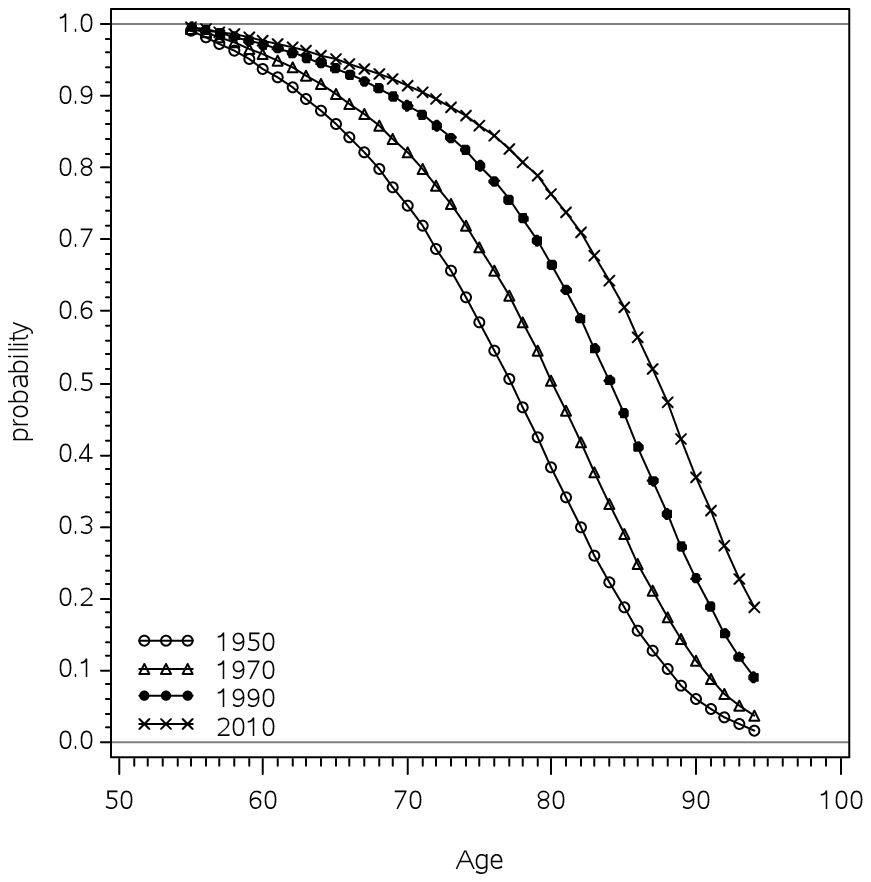}
}
\caption{Survival curves for French females, $x_0=55$}
\label{fig01}
\end{figure}

Let us see computations of survival probabilities using observed central death rates.
Estimates of $r_{x,t}$ and $S(x,t)$, taking $x_0=55$, for French females for some ages and years, 
calculated using historical data of $m_{x,t}$ obtained from Human Mortality Database \cite{HumanMortalityDatabase2013}, are presented in Figure \ref{fig01}.
In both panels of this figure there are sharp shifts of the curves towards the right.
This is an evidence of mortality improvements over years.
However, these movements do not happen always in the same way.
The sequence of the curves of deaths reveals that these curves are increasingly narrower and thus their heights tend to be higher.
Moreover, the life span for French females shows no sign of approaching towards a fixed limit, 
since the last survival probability of the curve of deaths presented in Figure \ref{fig01} tends to increase when time increases.
All of this shows that the dynamics of $r_{x,t}$ and $S(x,t)$ are complex and their examinations thus need strategies to appropiately dissect their operations.

\begin{figure}[!ht]
\centering
\includegraphics[scale=0.70]{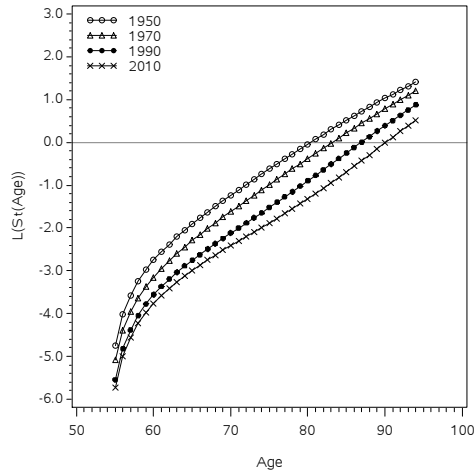}
\caption{Application of $L$ to survival functions for French females, $x_0=55$}
\label{fig02}
\end{figure}

In this paper we will use the transform $L$ to represent mortality dynamics.
To this aim, let us first see how this transform works when it is applied to survival functions, using the example introduced above.
Figure \ref{fig02} shows the results of the application of $L$ to the survival functions exhibited in Figure \ref{fig01}.
The resulting curves are quite smooth for most of ages with respect to the survival functions shown in Figure \ref{fig01}, excepting for ages near $x_0$.
Showing initially a concave shape, it seems, for the age period 70 - 94, that the transformed curves become right linear when time increases.
Moreover, they present downward shifts between the years considered, but not in a constant way and also not holding always the same curve shapes.
Furthermore, when age approaches towards the lowest or the highest studied ages then the curves are increasingly close each other.

\begin{figure}[!ht]
\centering
\includegraphics[scale=0.70]{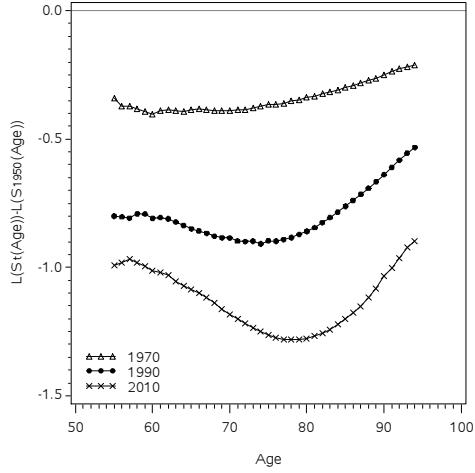}
\caption{Differences of $L(S(x))$ with respect to 1950 for French females, $x_0=55$}
\label{fig03}
\end{figure}

Next, we formulate relationships like \eqref{eq:20150117:001} or \eqref{eq:20150117:002} in terms of $L$.
We consider the difference of transformations of survival functions computed for years $t$ and $t_0$ where $t_0$ is a given year, i.e. $L(S_t(x))-L(S_{t_0}(x))$.
Using the data introduced above, Figure \ref{fig03} shows the outputs of these computations when the year of reference is 1950.
We now have only three curves since the curve corresponding to 1950 coincides with the horizontal axis and it then is not useful in our analysis.
They are well-distinguished and well-separated among them with respect to those exhibited in Figure \ref{fig02}, and show concave shapes increasingly more pronounced and downward shifts when time increases.
These curves do not have unstable behaviors at high ages, i.e. near 94 years old.
It is in advantage to model those ends of curves because the structure to describe is simple.
Furthermore, the transformation $L(S_t(x))-L(S_{t_0}(x))$ seems a promising method to study very high ages, e.g. from 90 and over, since the trend of its outcomes seems to be linear.

\section{New stochastic mortality models based on $L$}
\label{model}

\subsection{New model}

We aim to model curves like those presented in Figure \ref{fig03}.
For this purpose, given the non-linear nature of the differences of $L(S(x))$, we borrow representations of mortality like those of James and Segal in \cite{JamesSegal1982} or Cairns et al. in \cite{CairnsBlakeDowd2006} and use them to describe such differences.
The structures of the methods of these authors are ease to implement and to interpret.
In this paper we adopt the representation given by Cairns et al. and allow greater flexibility to age-parameters.
Our model, which we called the SL model, for representing and forecasting mortality is, given $x_0$, $x_{\min}=x_0$, $x_{\max}$, $t_0$, and $t_{\min}=t_0+1$,
\begin{equation}\label{eq:20150118:001}
L(S_t(x))-L(S_{t_0}(x))=\alpha_{1,t}+\alpha_{2,t}\kappa_x\textrm{,\quad $x_{\min}\leq x\leq x_{\max}$, $t\geq t_{\min}$,}
\end{equation}
where $\kappa_x$ are age-specific constants and, $\alpha_{1,t}$ and $\alpha_{2,t}$ are stochastic processes that are assumed to be measurable at time $t$.
Note that $\kappa_x$ varies linearly in \cite{CairnsBlakeDowd2006}, but in \eqref{eq:20150118:001} it could vary non-linearly.

To project $L(S_t(x))-L(S_{t_0}(x))$, we adopt for the two-dimensional time series $\boldsymbol{\alpha}_t = (\alpha_{1,t},\alpha_{2,t})'$ the dynamics given for the two-dimensional stochastic parameter vector from the CBD model \cite{CairnsBlakeDowd2006}, i.e. $\boldsymbol{\alpha}_t$ is modeled by the following two-dimensional random walk with drift:
\begin{equation}\label{eq:20150118:002}
\boldsymbol{\alpha}_{t+1} = \boldsymbol{\alpha}_t + \mathbf{a} + \mathbf{A}\mathbf{Z}_{t+1}\textrm{,}
\end{equation}
where $\mathbf{a}$ is a constant $2\times1$ vector, $\mathbf{A}$ is a deterministic $2\times2$ upper triangular matrix, and $\mathbf{Z}_t$ is a two-dimensional standard normal random variable.

Note that from projections of $L(S_t(x))-L(S_{t_0}(x))$ one can easily compute any mortality variable using the relationships given in Section \ref{examination}, say for instance survival probabilities, $q_{x,t}$, and $m_{x,t}$.


Next, vaying $t_0$ we get extensions of \eqref{eq:20150118:001}, i.e. $t_0$ can be considered as another parameter more in \eqref{eq:20150118:001} and also $x_0$.
Note that, fixing $t_{\min}$, $t_0$ can be chosen smaller than $t_{\min}-1$.
Some of these features will be analyzed and exploited in the applications to be presented later.

\begin{figure}[!ht]
\centering
\subfiguretopcapfalse
\subfigure[1952]{
\includegraphics[scale=0.35]{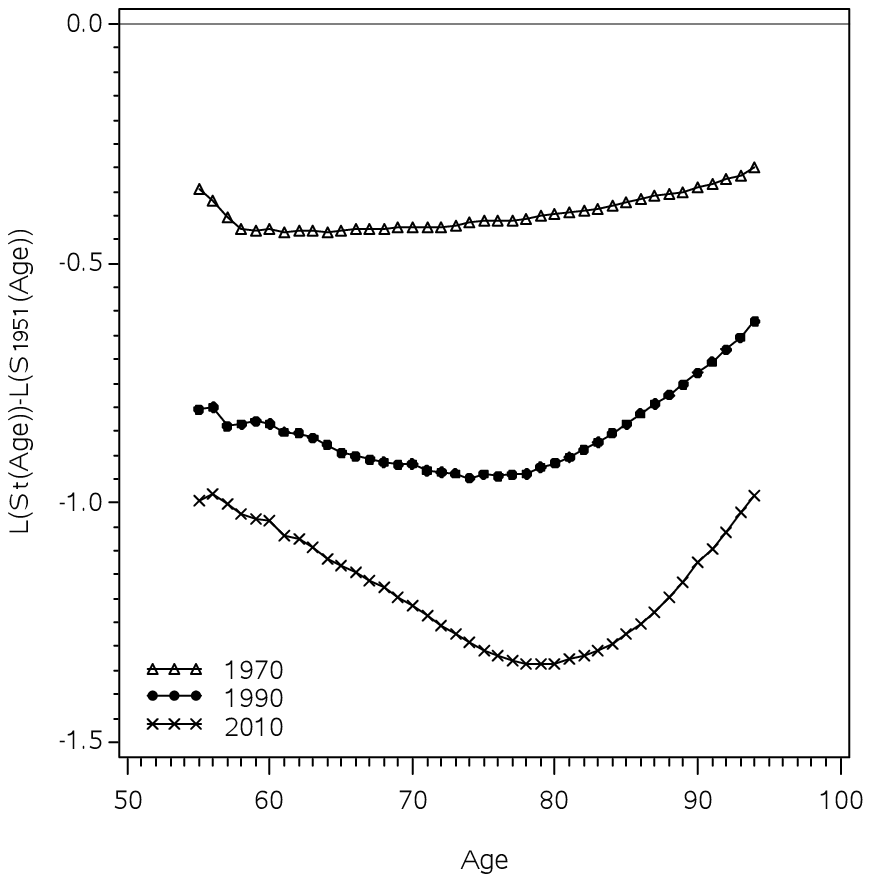}
}
\hspace{1mm}
\subfigure[1953]{
\includegraphics[scale=0.35]{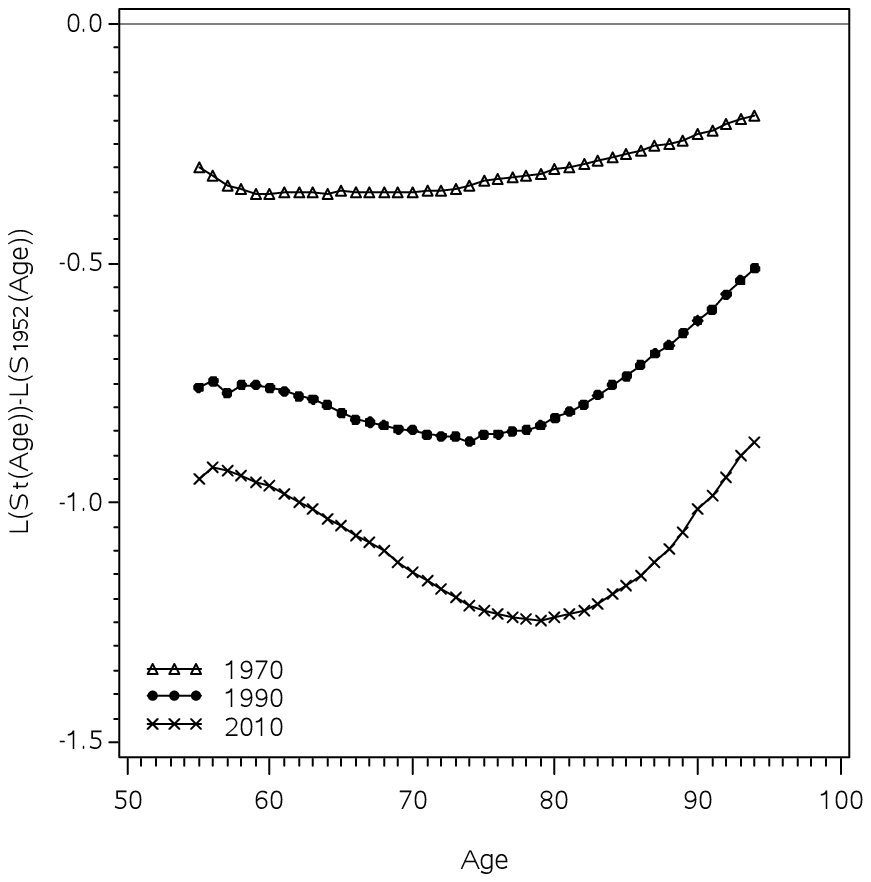}
}
\hspace{1mm}
\subfigure[1954]{
\includegraphics[scale=0.35]{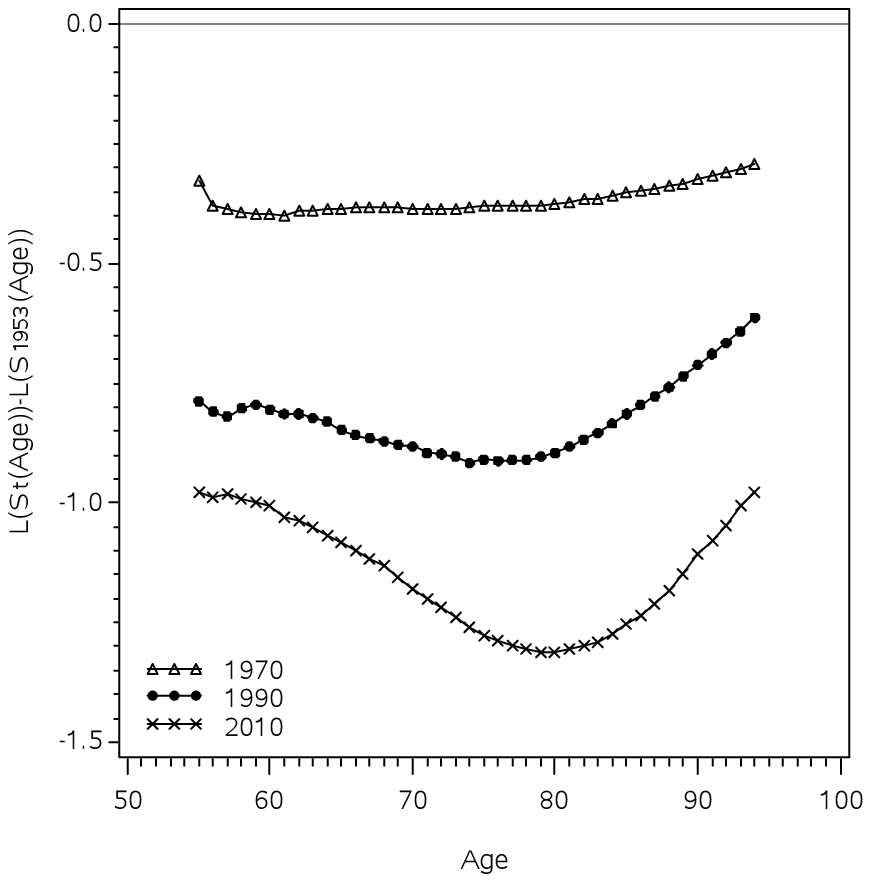}
}
\hspace{1mm}
\subfigure[1955]{
\includegraphics[scale=0.35]{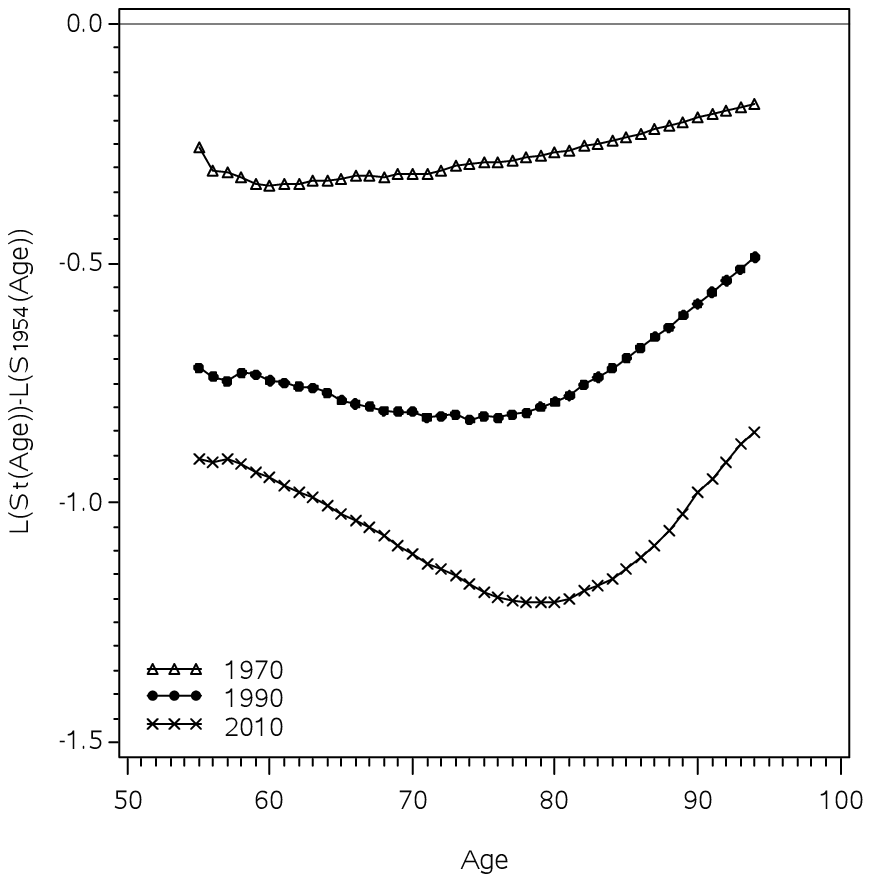}
}
\caption{Differences of $L(S(x))$ with respect to several years for French females, $x_0=55$}
\label{fig04}
\end{figure}

Let us see numerical illustrations of $L(S_t(x))-L(S_{t_0}(x))$ when $t_0$ varies.
Figure \ref{fig04} shows curves like those exhibited in Figure \ref{fig03} for French females but now for $t_0$ = 1951, 1952, 1953, and 1954.
Note that if $t_0$ reaches 1955, then the curve corresponding to $t=1955$ does not have interest in the analysis.
These new curves are lightly different among them, and it seems that all of them move upward when $t_0$ increases.

\begin{figure}[!ht]
\centering
\subfiguretopcapfalse
\subfigure[1952]{
\includegraphics[scale=0.35]{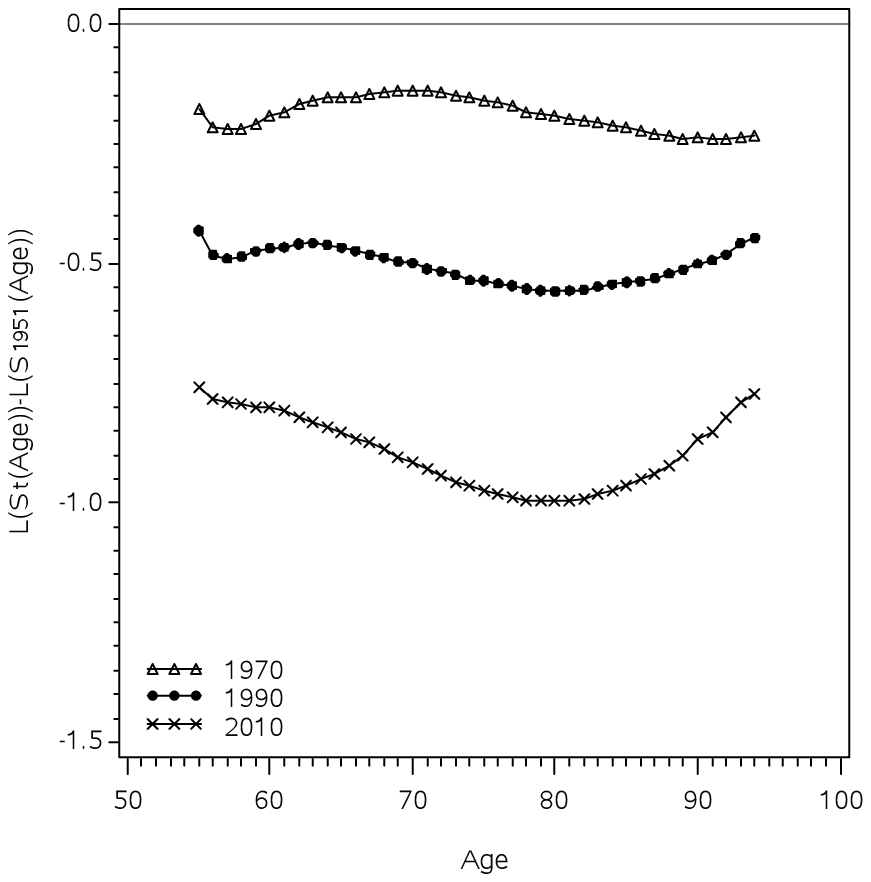}
}
\hspace{1mm}
\subfigure[1953]{
\includegraphics[scale=0.35]{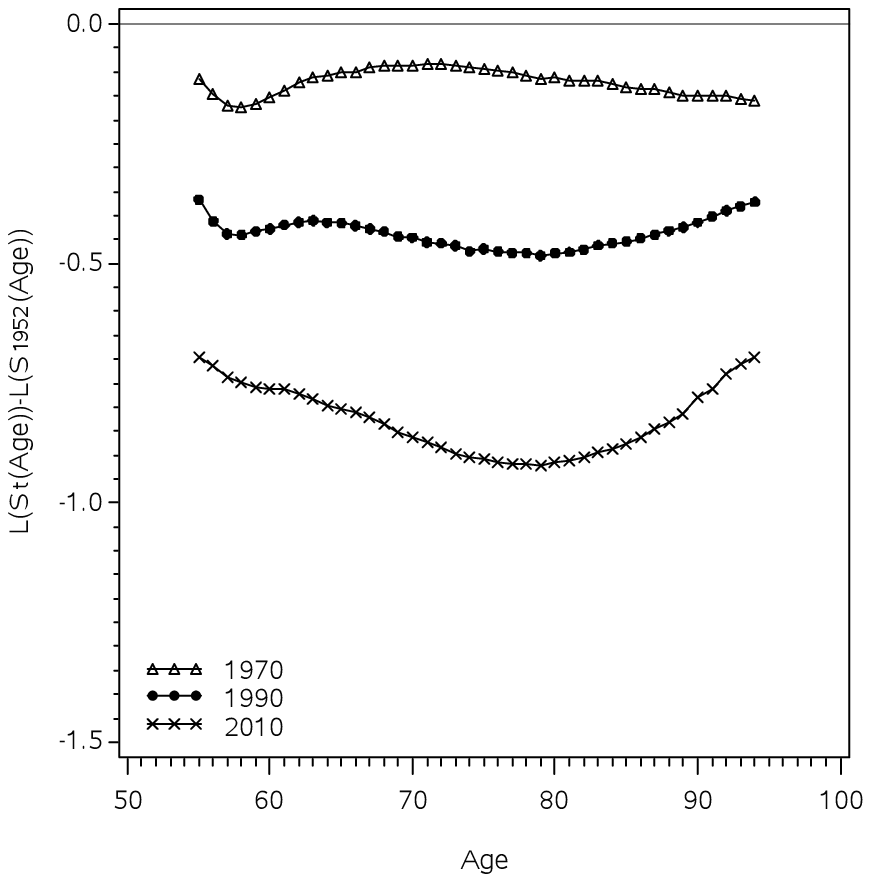}
}
\hspace{1mm}
\subfigure[1954]{
\includegraphics[scale=0.35]{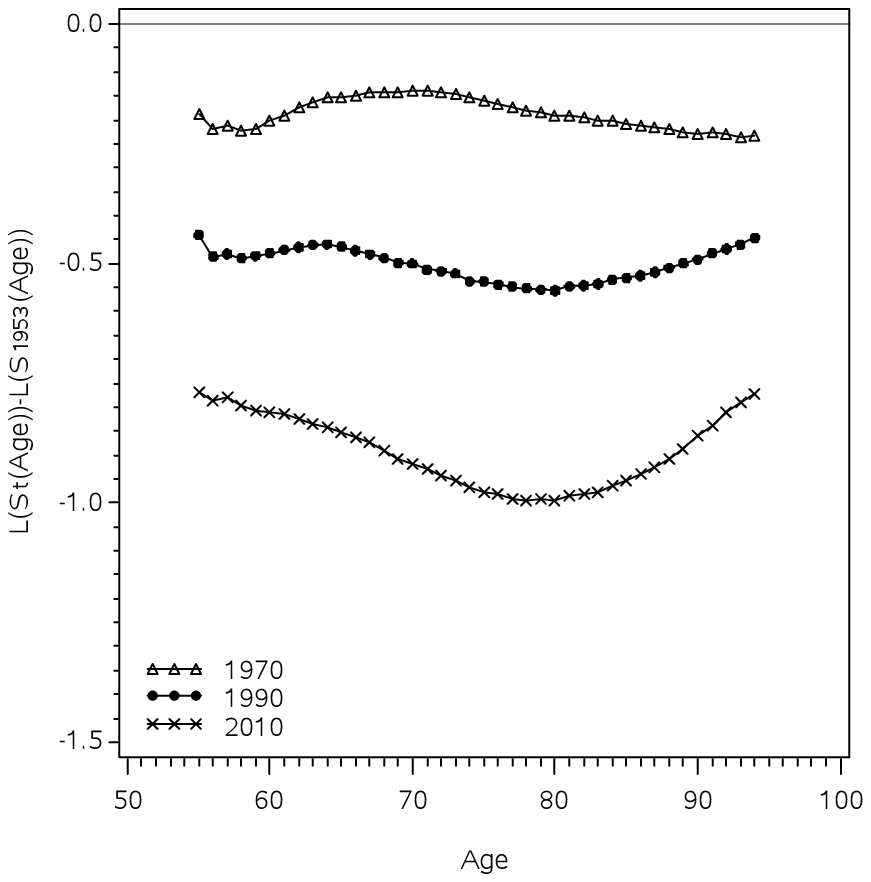}
}
\hspace{1mm}
\subfigure[1955]{
\includegraphics[scale=0.35]{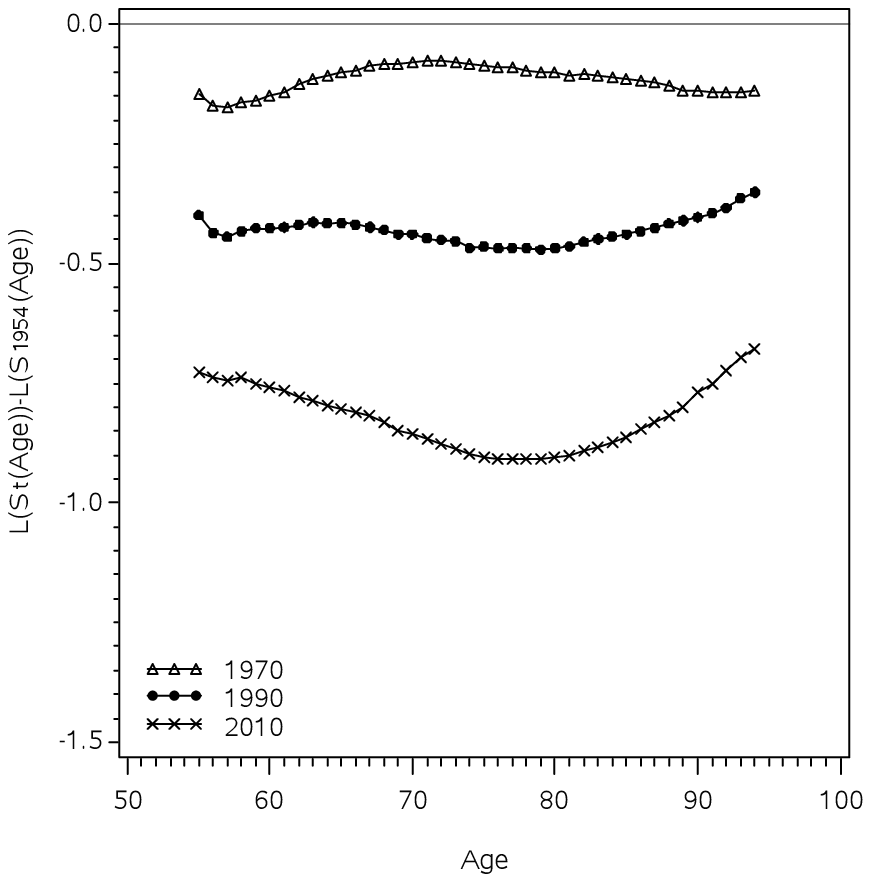}
}
\caption{Differences of $L(S(x))$ with respect to several years for French males, $x_0=55$}
\label{fig05}
\end{figure}

The corresponding curves of Figure \ref{fig04} for French males are presented in Figure \ref{fig05}.
They are a bit different from the French female curves.
Now the lower curves hold concave shapes, but the upper curves are not always concave and present convex pieces.

\subsection{OLS estimation}

The model \eqref{eq:20150118:001}, given $t_0$ and $x_0$, is fitted to computed values of $L(S_t(x))-L(S_{t_0}(x))$ using ordinary least squares (OLS).
It consists in to minimize
\begin{equation}\label{eq:20150118:003}
\sum_{x,t}\,\left(L\left({S}_t(x)\right)-L\left({S}_{t_0}(x)\right)-\alpha_{1,t}-\alpha_{2,t}\kappa_x\right)^2\textrm{.}
\end{equation}
As noticed by Brouhns et al. \cite{BrouhnsDenuitVermunt2002}, models like \eqref{eq:20150118:001} are not simple regression models since there are no observed covariates in the right-hand side.
Hence alternative ways to find values of $\alpha_{1,t}$, $\alpha_{2,t}$, and $\kappa_x$ minimizing \eqref{eq:20150118:003} are required.
We notice that the uni-dimensional or elementary Newton method proposed by Goodman \cite{Goodman1979} can be used in this case.
This method works like the iterative Newton method but considering in each iteration one parameter at a time.
In this way each parameter is updated in the iteration $k$ before the iteration of the next parameter in the same iteration $k$.
Iteration moves from $k$ to $k+1$ when all the parameters have iterated in the iteration $k$.
This process continues until to find the convergence of the sequence of parameter estimates or another stopping condition.

The fitting of \eqref{eq:20150118:001} to data showed that some parameters may require slow convergences.
It was implemented by reducing the variations among iterations, for instance multiplying such variations by a factor $\gamma<1$.
This is taken into account in the following iteration algorithm designed to minimize \eqref{eq:20150118:003}:

\begin{enumerate}
\item
Initializations:

\begin{enumerate}
\item
Initialization of $\alpha_{1,t}$, $\alpha_{2,t}$, and $\kappa_x$: $\hat{\alpha}_{1,t}^{(0)}$, $\hat{\alpha}_{2,t}^{(0)}$, and $\hat{\kappa}_x^{(0)}$.

A strategy to obtain these initial values is to fix $\kappa_x$ using a known function of $x$.
Then one can find estimates $\hat{\alpha}_{1,t}$ and $\hat{\alpha}_{2,t}$ applying linear regressions of $L\big({S}_t(x)\big)-L\big({S}_{t_0}(x)\big)$ on $\kappa_x$ for each $t$.

\item
Set the maximal number of iterations $k_{\max}$.

\item
Set $k=1$.

\end{enumerate}

\item
While not convergent and $k<k_{\max}$ do:

\begin{enumerate}
\item
For $k$ do:

\begin{quote}
$
\displaystyle
\hat{\alpha}_{1,t}^{(k)}=\hat{\alpha}_{1,t}^{(k-1)}+\gamma\sum_t\left(L\left({S}_t(x)\right)-L\left({S}_{t_0}(x)\right)-\hat{\alpha}_{1,t}^{(k-1)}-\hat{\alpha}_{2,t}^{(k-1)}\hat{\kappa}_{x}^{(k-1)}\right)
$;
\end{quote}

\begin{quote}
$
\displaystyle
\hat{\alpha}_{2,t}^{(k)}=\hat{\alpha}_{2,t}^{(k-1)}+\gamma
\frac{\sum_t\hat{\kappa}_{x}^{(k-1)}\left(L\left({S}_t(x)\right)-L\left({S}_{t_0}(x)\right)-\hat{\alpha}_{1,t}^{(k)}-\hat{\alpha}_{2,t}^{(k-1)}\hat{\kappa}_{x}^{(k-1)}\right)}
{\sum_t\left(\hat{\kappa}_{x}^{(k-1)}\right)^2}
$; and,
\end{quote}

\begin{quote}
$
\displaystyle
\hat{\kappa}_{x}^{(k)}=\hat{\kappa}_{x}^{(k-1)}+\gamma
\frac{\sum_x\hat{\alpha}_{2,t}^{(k)}\left(L\left({S}_t(x)\right)-L\left({S}_{t_0}(x)\right)-\hat{\alpha}_{1,t}^{(k)}-\hat{\alpha}_{2,t}^{(k)}\hat{\kappa}_{x}^{(k-1)}\right)}
{\sum_x\left(\hat{\alpha}_{2,t}^{(k)}\right)^2}
$.
\end{quote}

\item
$k=k+1$.

\end{enumerate}

\end{enumerate}

The criterion of convergence is to reach $\big|\hat{v}^{(k)}-\hat{v}^{(k-1)}\big|<\epsilon$, being $v$ any of $\alpha_{1,t}$, $\alpha_{2,t}$, or $\kappa_{x}$, for a reasonably small value of $\epsilon$.
When this stopping criterion is satisfied, the OLS estimates of $\alpha_{1,t}$, $\alpha_{2,t}$, and $\kappa_{x}$ using Newton's method, $\hat{\alpha}_{1,t}$, $\hat{\alpha}_{2,t}$, and $\hat{\kappa}_{x}$, are the final iterative estimates $\hat{\alpha}^{(k)}_{1,t}$, $\hat{\alpha}^{(k)}_{2,t}$, and $\hat{\kappa}^{(k)}_{x}$.
In applications to be presented later $\epsilon=10^{-8}$.
On $k_{\max}$, the maximal number of iterations, we will fix it in 5,000.

\section{Numerical illustrations}
\label{illustrations}

At present we apply the LS model to mortality data of seven industrialized countries and compare their mortality projections with those of the LC and CBD models.
We will examine $q_{x,t}$ at higher ages (60 - 94) using in-sample and out-of-sample goodness-of-fit criteria and considering some age periods.

To this aim, we start with the description of data to be considered and the presentation of the information to be modeled.
Next, the LC and CBD models are briefly described.
In the last part the results are shown and discussed.

\subsection{Data}

In this paper we use sex-based mortality data from populations of Belgium (BE), France (FR), Italy (IT), Japan (JP), Sweden (SW), United Kingdom (UK), and United States (US).
Data, obtained from Human Mortality Database (HMD) \cite{HumanMortalityDatabase2013}, correspond to ages from 60 to 94, and to years from 1960 to 2009.
No adjustment is made to the data.

A typical dataset consists of the observed variable ${D}_{x,t}$ and the computed variable ${E}_{x,t}$ over the above-mentioned ranges of ages and years.
${D}_{x,t}$ is the number of deaths during calendar year $t$ aged $x$ last birthday and ${E}_{x,t}$ represents an estimated average, during calendar year $t$, of the number of people alive who were aged $x$ last birthday.

With these variables $m_{x,t}$ is estimated by (see e.g. \cite{protocol})
$$
\hat{m}_{x,t}=\frac{{D}_{x,t}}{{E}_{x,t}}\textrm{,}
$$
and $q_{x,t}$ and $S_t(x)$ can then be computed using the relationships presented in previous Section.
Recall that $S_t(x)$ is involved in the LS model, and we will see that $m_{x,t}$ and $q_{x,t}$ are taken into account in the LC and CBD models.

\subsection{Benchmark models}

We compare mortality projections of the SL model with those of the LC and CBD models.
These two last models are taken as benchmark models since they are well-known and extensively used.
Brief descriptions of these models follow.

\subsubsection{Lee-Carter model}

This model proposed by Lee and Carter in \cite{LeeCarter1992} is given by

\begin{equation}\label{eq:intro:lc::01}
\log\left(m_{x,t}\right)=\alpha_x+\beta_x\kappa_t\textrm{,}
\end{equation}

where $\alpha_x$ and $\beta_x$ are specific constants at age $x$, and $\kappa_t$ is a 
time-varying index.

Since the parameter estimation for the LC model is not unique, the following two constraints are adopted:
$$
\sum_t\kappa_t=0\textrm{\quad and\quad}\sum_x\beta_x=1\textrm{.}
$$
For projecting mortality the resulting estimate of the parameter $\kappa_t$ is projected as a stochastic time series
using standard Box-Jenkins methods.

To apply this model in our research we use the function \texttt{LCA} provided by the library \texttt{Demography} implemented in the language \texttt{R} (see \cite{HyndmanBoothTickleMaindonald2015}).
This function estimates the parameters $\alpha_x$, $\beta_x$, and $\kappa_t$ using singular value decomposition, following the methodology proposed by Lee and Carter.

\subsubsection{Cairns-Blake-Dowd model}

This two-factor stochastic mortality model was introduced by \cite{CairnsBlakeDowd2006}.
It is given at age $x$, $x\in[x_1,x_2]$, and time $t$ by

\begin{equation}\label{eq:intro:cbd::01}
\textrm{logit}\left(q_{x,t}\right)=\kappa_{1,t}+\kappa_{2,t}\,(x-\bar{x})\textrm{,}
\end{equation}

where $\kappa_{1,t}$ and $\kappa_{2,t}$ are stochastic processes that are assumed to be measurable at time $t$, and $\bar{x}$ is the age mean of the range of ages analyzed.
The two dimensional time-series $\boldsymbol{\kappa}_t=(\kappa_{1,t},\kappa_{2,t})'$ is modeled as in \eqref{eq:20150118:002}.

Following \cite{CairnsBlakeDowd2006}, for each $t$, $\kappa_{1,t}$ and $\kappa_{2,t}$ are estimated using least squares by transforming $q_x$ to 
$\textrm{logit}\left(q_{x,t}\right)=\kappa_{1,t}+\kappa_{2,t}\,(x-\bar{x}) +\textrm{error}$.

\subsection{Analysis of results}

In this subsection we compare the in-sample and out-of-sample predictive abilities of the LS, LC, and CBD models with respect to data of sex-based $q_{x,t}$ of seven countries.
We examine this mortality variable since it is commonly used in practice, for instance to build mortality tables and to compute annuities.

We focus on mortality at higher ages because it is related to population ageing.
Age periods from $x_{\min}$ to $x_{\max}$ are analyzed taking $x_{\min}$ = 60, 65 and $x_{\max}$ = 89, 94.

Observed mortality from 1960 to 2009 is used to evaluate in-sample and out-of-sample mortality projection accuracies.
This year period is divided in two, the first, from $t_{\min}$ = 1960 to $t_{\max}$ = 1989, to fit the models to data and the second, from 1990 to 2009, to forecast mortality.
For the LS model we fix $t_0=t_{\min}-1$ in \eqref{eq:20150118:001}.

We examine both in-sample and out-of-sample projection accuracies of $q_{x,t}$ employing a couple of goodness-of-fit criteria often used to evaluate projection performance (see e.g. \cite{LeeMiller2001} and \cite{ChenLiaoShangLi2009}).
The first criterion is the mean squared error (MSE) defined as
$$
\textrm{MSE}=\frac{1}{n}\sum_{i=1}^{n}\big(X_i-\hat{X}_i\big)^2\textrm{}
$$
where $X_i$ and $\hat{X}_i$ are the observed and estimated values of mortality in fitting or forecasting samples,
and the second criterion is the mean absolute percentage error (MAPE) defined as
$$
\textrm{MAPE}=\frac{1}{n}\sum_{i=1}^{n}\frac{\big|X_i-\hat{X}_i\big|}{\hat{X}_i}\times100~\%\textrm{.}
$$
A smaller MSE or MAPE value indicates a better fit to the data on a given period.
However, the lowest values of these indexes are not associated to the same model necessarily.
We will come back on this issue later.

For each combination of the values of $x_{\min}$ and $x_{\max}$ mentioned above, we present values of MSE and MAPE in Tables 
\ref{tab01_60196011_1}, \ref{tab01_65196011_1}, \ref{tab01_60196011_2}, and \ref{tab01_65196011_2}.
Each of these tables displays countries in rows and sex categories in columns.
For each country, the CBD, LC, and LS models are nested.
Values of MSE and MAPE are shown for fitting and forecasting year periods, nested in each category of sex. 
Additionally, in each combination of country and sex and fitting / forecasting year period, the lowest values of MSE and MAPE among the models studied are highlighted, which allow the identification of the models with the best fits.
Besides, the values of MSE are displayed with two decimals and those of MAPE with one decimal, 
this may produce some confusion if a highlighted value equals other values, but there is no conflict since such selection was done using more decimals than those displayed.

A general finding through all these tables is that, as expected, the values of MSE or MAPE on forecasting periods are always much higher than their corresponding values on fitting periods.
Also, as indicated above, for fitting as well as for forecasting periods, not always the model with the highest value of MSE is the model with the highest value of MAPE.
Additionally, the distributions of the highlighted models are frequently different from a table to another, but in a few cases, for a pair of tables and a given country, one can find the same selection of models.
Another finding is that the values of MSE or MAPE, over fitting or forecasting periods, are in general higher for males than for females, 
which suggests that the modelization of mortality is more difficult for males than for females.

On the fitting period, for females, the highlighted values are concentrated mainly on the LC model and a few cases on the LS model, and none of these values is associated to the CBD model.
An interesting feature of the LC model is that it is highlighted almost everywhere when $x_{\max}$ is 94.
Regarding males, the highlighted models are concentrated mainly on the CBD and LC models and a few cases on the LS model.
Moreover, no model is predominant through all values of $x_{\min}$ and $x_{\max}$.

On the forecasting period, a first observation is that not always the best model over the fitting period corresponds to the best model over the forecasting period.
This implies that the distribution of the highlighted values varies between these two periods.
On females, we have a sharp presence of the SL model, excepting the case $x_{\min}$ = 60 and $x_{\max}$ = 94.
It performs better than the benchmark models for US and UK, and gives the best mortality projection accuracies for JP and SW when $x_{\min}$ = 65 and $x_{\max}$ is 89 or 94.
Highlighted values associated to the LC and CDB models appear in a few times.
On males, there is a strong presence of the CBD model for an important number of countries through all values of $x_{\min}$ and $x_{\max}$, mainly BE, NE, SW, and US.
The LC and SL models are found in a few cases.
Nevertheless the scarce selection of the SL model as one of the best models for males, it gives the best mortality projection accuracies for UK for any value of $x_{\min}$ or $x_{\max}$.

\begin{table}
{\small
\begin{center}
\begin{tabular}{llrrcrrcrrcrr}
\hline
 & & \multicolumn{5}{c}{Female} & & \multicolumn{5}{c}{Male} \\
\cline{3-7}\cline{9-13}
Country & Model & \multicolumn{2}{c}{Fitt. Period} & & \multicolumn{2}{c}{Frcs. Period} & & \multicolumn{2}{c}{Fitt. Period} & & \multicolumn{2}{c}{Frcs. Period} \\
\cline{3-4}\cline{6-7}\cline{9-10}\cline{12-13}
 & & \makebox[0.55cm][c]{\centering MSE$^*$} & \makebox[0.55cm][c]{\centering MAPE} & & \makebox[0.55cm][c]{\centering MSE$^*$} & \makebox[0.55cm][c]{\centering MAPE} & & \makebox[0.55cm][c]{\centering MSE$^*$} & \makebox[0.55cm][c]{\centering MAPE} & & \makebox[0.55cm][c]{\centering MSE$^*$} & \makebox[0.55cm][c]{\centering MAPE} \\
\hline
BE  & CBD  & \cellcolor[gray]{1.0} 0.06  & \cellcolor[gray]{1.0} 3.2  & & \cellcolor[gray]{1.0} 0.12  & \cellcolor[gray]{1.0} 7.8  & & \cellcolor[gray]{0.8} 0.17  & \cellcolor[gray]{0.8} 3.0  & & \cellcolor[gray]{1.0} 1.02  & \cellcolor[gray]{0.8} 11.3  \\
 & LC  & \cellcolor[gray]{1.0} 0.06  & \cellcolor[gray]{0.8} 2.5  & & \cellcolor[gray]{0.8} 0.09  & \cellcolor[gray]{0.8} 4.9  & & \cellcolor[gray]{1.0} 0.19  & \cellcolor[gray]{1.0} 3.0  & & \cellcolor[gray]{0.8} 1.0  & \cellcolor[gray]{1.0} 11.8  \\
 & SL  & \cellcolor[gray]{0.8} 0.05  & \cellcolor[gray]{1.0} 2.7  & & \cellcolor[gray]{1.0} 0.15  & \cellcolor[gray]{1.0} 6.5  & & \cellcolor[gray]{1.0} 0.40  & \cellcolor[gray]{1.0} 3.5  & & \cellcolor[gray]{1.0} 1.2  & \cellcolor[gray]{1.0} 13.5  \\
\hline
FR  & CBD  & \cellcolor[gray]{1.0} 0.06  & \cellcolor[gray]{1.0} 4.2  & & \cellcolor[gray]{1.0} 0.14  & \cellcolor[gray]{1.0} 9.9  & & \cellcolor[gray]{1.0} 0.10  & \cellcolor[gray]{1.0} 3.1  & & \cellcolor[gray]{0.8} 0.35  & \cellcolor[gray]{1.0} 8.7  \\
 & LC  & \cellcolor[gray]{1.0} 0.02  & \cellcolor[gray]{1.0} 1.9  & & \cellcolor[gray]{1.0} 0.11  & \cellcolor[gray]{1.0} 7.8  & & \cellcolor[gray]{0.8} 0.09  & \cellcolor[gray]{0.8} 2.2  & & \cellcolor[gray]{1.0} 0.4  & \cellcolor[gray]{0.8} 7.2  \\
 & SL  & \cellcolor[gray]{0.8} 0.02  & \cellcolor[gray]{0.8} 1.8  & & \cellcolor[gray]{0.8} 0.09  & \cellcolor[gray]{0.8} 5.8  & & \cellcolor[gray]{1.0} 0.13  & \cellcolor[gray]{1.0} 2.5  & & \cellcolor[gray]{1.0} 0.4  & \cellcolor[gray]{1.0} 8.4  \\
\hline
IT  & CBD  & \cellcolor[gray]{1.0} 0.05  & \cellcolor[gray]{1.0} 3.1  & & \cellcolor[gray]{0.8} 0.11  & \cellcolor[gray]{1.0} 6.7  & & \cellcolor[gray]{0.8} 0.08  & \cellcolor[gray]{1.0} 2.6  & & \cellcolor[gray]{0.8} 0.42  & \cellcolor[gray]{1.0} 12.1  \\
 & LC  & \cellcolor[gray]{1.0} 0.04  & \cellcolor[gray]{0.8} 2.1  & & \cellcolor[gray]{1.0} 0.11  & \cellcolor[gray]{0.8} 4.6  & & \cellcolor[gray]{1.0} 0.10  & \cellcolor[gray]{1.0} 2.4  & & \cellcolor[gray]{1.0} 0.6  & \cellcolor[gray]{0.8} 11.9  \\
 & SL  & \cellcolor[gray]{0.8} 0.04  & \cellcolor[gray]{1.0} 2.1  & & \cellcolor[gray]{1.0} 0.13  & \cellcolor[gray]{1.0} 6.0  & & \cellcolor[gray]{1.0} 0.13  & \cellcolor[gray]{0.8} 2.4  & & \cellcolor[gray]{1.0} 0.4  & \cellcolor[gray]{1.0} 12.6  \\
\hline
JP  & CBD  & \cellcolor[gray]{1.0} 0.06  & \cellcolor[gray]{1.0} 3.1  & & \cellcolor[gray]{0.8} 0.07  & \cellcolor[gray]{1.0} 11.3  & & \cellcolor[gray]{1.0} 0.09  & \cellcolor[gray]{1.0} 2.1  & & \cellcolor[gray]{1.0} 0.13  & \cellcolor[gray]{0.8} 9.7  \\
 & LC  & \cellcolor[gray]{0.8} 0.04  & \cellcolor[gray]{1.0} 1.9  & & \cellcolor[gray]{1.0} 0.28  & \cellcolor[gray]{1.0} 13.9  & & \cellcolor[gray]{0.8} 0.06  & \cellcolor[gray]{0.8} 1.8  & & \cellcolor[gray]{0.8} 0.1  & \cellcolor[gray]{1.0} 11.3  \\
 & SL  & \cellcolor[gray]{1.0} 0.05  & \cellcolor[gray]{0.8} 1.9  & & \cellcolor[gray]{1.0} 0.07  & \cellcolor[gray]{0.8} 9.0  & & \cellcolor[gray]{1.0} 0.09  & \cellcolor[gray]{1.0} 2.0  & & \cellcolor[gray]{1.0} 0.6  & \cellcolor[gray]{1.0} 10.1  \\
\hline
SW  & CBD  & \cellcolor[gray]{1.0} 0.09  & \cellcolor[gray]{1.0} 3.8  & & \cellcolor[gray]{1.0} 0.46  & \cellcolor[gray]{1.0} 10.2  & & \cellcolor[gray]{0.8} 0.08  & \cellcolor[gray]{0.8} 2.4  & & \cellcolor[gray]{0.8} 0.30  & \cellcolor[gray]{0.8} 9.8  \\
 & LC  & \cellcolor[gray]{0.8} 0.05  & \cellcolor[gray]{0.8} 2.8  & & \cellcolor[gray]{0.8} 0.19  & \cellcolor[gray]{0.8} 8.0  & & \cellcolor[gray]{1.0} 0.11  & \cellcolor[gray]{1.0} 2.4  & & \cellcolor[gray]{1.0} 0.4  & \cellcolor[gray]{1.0} 10.4  \\
 & SL  & \cellcolor[gray]{1.0} 0.11  & \cellcolor[gray]{1.0} 3.3  & & \cellcolor[gray]{1.0} 0.32  & \cellcolor[gray]{1.0} 8.6  & & \cellcolor[gray]{1.0} 0.23  & \cellcolor[gray]{1.0} 3.0  & & \cellcolor[gray]{1.0} 0.4  & \cellcolor[gray]{1.0} 10.7  \\
\hline
UK  & CBD  & \cellcolor[gray]{1.0} 0.04  & \cellcolor[gray]{1.0} 2.5  & & \cellcolor[gray]{1.0} 0.16  & \cellcolor[gray]{1.0} 10.4  & & \cellcolor[gray]{0.8} 0.07  & \cellcolor[gray]{1.0} 2.1  & & \cellcolor[gray]{1.0} 1.74  & \cellcolor[gray]{1.0} 15.3  \\
 & LC  & \cellcolor[gray]{0.8} 0.02  & \cellcolor[gray]{0.8} 1.5  & & \cellcolor[gray]{1.0} 0.11  & \cellcolor[gray]{1.0} 9.7  & & \cellcolor[gray]{1.0} 0.08  & \cellcolor[gray]{0.8} 1.9  & & \cellcolor[gray]{1.0} 1.4  & \cellcolor[gray]{1.0} 15.2  \\
 & SL  & \cellcolor[gray]{1.0} 0.02  & \cellcolor[gray]{1.0} 1.8  & & \cellcolor[gray]{0.8} 0.08  & \cellcolor[gray]{0.8} 8.5  & & \cellcolor[gray]{1.0} 0.12  & \cellcolor[gray]{1.0} 2.1  & & \cellcolor[gray]{0.8} 0.9  & \cellcolor[gray]{0.8} 13.3  \\
\hline
US  & CBD  & \cellcolor[gray]{1.0} 0.07  & \cellcolor[gray]{1.0} 4.2  & & \cellcolor[gray]{1.0} 0.59  & \cellcolor[gray]{1.0} 6.8  & & \cellcolor[gray]{1.0} 0.05  & \cellcolor[gray]{1.0} 2.1  & & \cellcolor[gray]{0.8} 0.28  & \cellcolor[gray]{0.8} 7.3  \\
 & LC  & \cellcolor[gray]{0.8} 0.01  & \cellcolor[gray]{0.8} 1.6  & & \cellcolor[gray]{1.0} 0.31  & \cellcolor[gray]{1.0} 6.6  & & \cellcolor[gray]{0.8} 0.04  & \cellcolor[gray]{0.8} 1.7  & & \cellcolor[gray]{1.0} 0.3  & \cellcolor[gray]{1.0} 7.5  \\
 & SL  & \cellcolor[gray]{1.0} 0.02  & \cellcolor[gray]{1.0} 2.0  & & \cellcolor[gray]{0.8} 0.24  & \cellcolor[gray]{0.8} 5.9  & & \cellcolor[gray]{1.0} 0.06  & \cellcolor[gray]{1.0} 2.0  & & \cellcolor[gray]{1.0} 0.4  & \cellcolor[gray]{1.0} 7.9  \\
\hline
\multicolumn{12}{l}{\quad MSE$^*$ = 10,000 $\times$ MSE.}
\end{tabular}
\end{center}
}
\caption{\label{tab01_60196011_1} MSE and MAPE of $q_{x,t}$ for $x_{\min}$ = 60 and  $x_{\max}$ = 89}
\end{table}

The values of MSE and MAPE allow the identification of the best model among the LC, CBD, and LS models, but without any assessment on how well these models fit, i.e. is any of these models acceptable?
For answering this question, we focus, for a given age $x$, on the rate of mortality improvement (MI) over the forecasting year period.
This rate indicates how mortality rates change with respect to mortality rates of a particular group at a specific point in time.

Rates of MI are often applied to initial mortality levels to establish generational tables to obtain mortality estimates at any future point in time (see e.g. \cite{OECD}).
These rates are also analyzed in function of economical and medical factors to explore their dynamics (see e.g. \cite{RMS2012}).
In this paper rates of MI are used to assess the performance of the models.

The definition of the rate of MI is not unique (see e.g. \cite{AndreevVaupel2005} and \cite{SOA2011}).
These rates may be computed as arithmetic or logarithmic rates, they may be based on $q_{x,t}$ or on $m_{x,t}$, and they may be calculated with respect to a given year or to a previous year.
We compute annual logarithmic rates of MI with respect to 1989 by, for $x=65$ and considering $x_{\min}=65$ and $x_{\max}=94$, for $t>1989$,
$$
\Delta_{65,t}=-\log\left(\frac{q_{65,t}}{{q}_{65,1989}}\right)\times100\textrm{,}
$$
where ${q}_{65,1989}$ is the value of $q_{65,t}$ observed in 1989.
Note that this definition can be seen as the cumulative rate of MI in year $t$ with respect to 1989.
This rate is expected to be increasing over the years since in the last decades $q_{x,t}$ has shown downward trends.

\begin{figure}[!ht]
\centering
\includegraphics[scale=1.0]{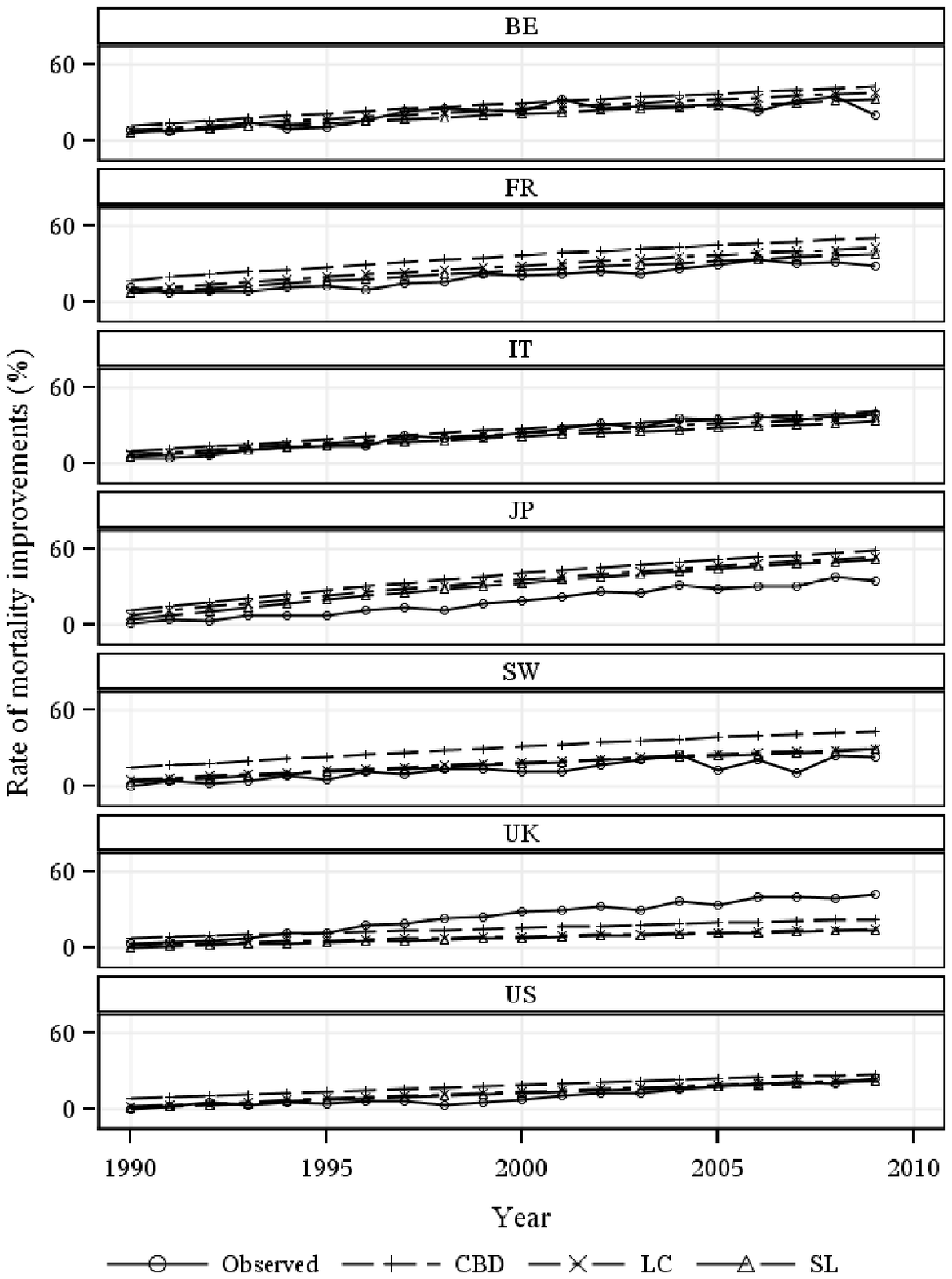}
\caption{Rates of MI with respect to 1989, taking $x_{\min}=65$ and $x_{\max}=89$,  females and $x=65$}
\label{fig11}
\end{figure}

\begin{figure}[!ht]
\centering
\includegraphics[scale=1.0]{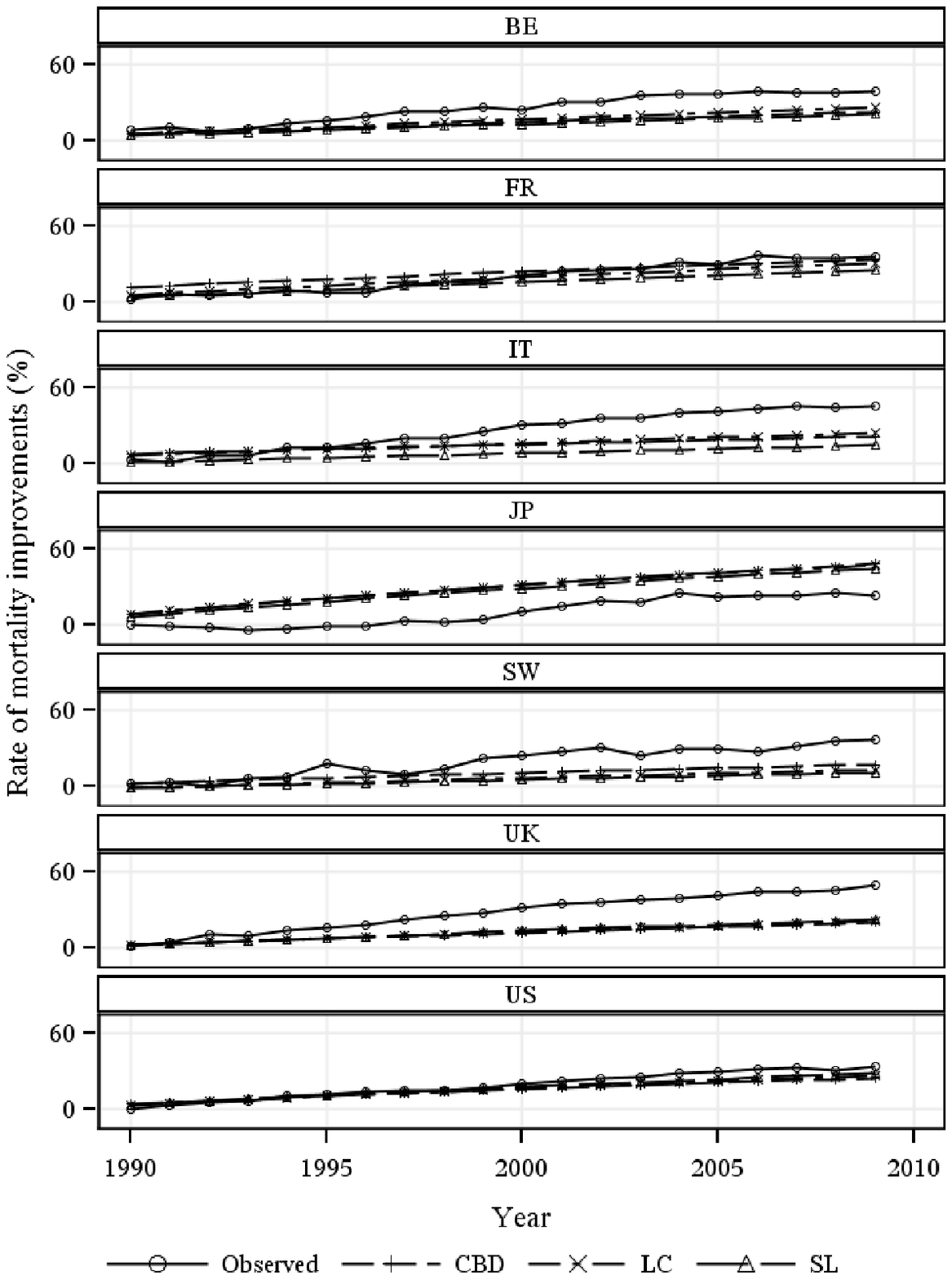}
\caption{Rates of MI with respect to 1989, taking $x_{\min}=65$ and $x_{\max}=89$,  males and $x=65$}
\label{fig12}
\end{figure}

Figure \ref{fig11} shows the observed and projected rates of MI obtained for females and for each country.
The corresponding rates for males are presented in Figure \ref{fig12}.
Through all these figures the curves have upward trends with variations from one year to the next, which is an empirical evidence of the uncertainty of the rates involved.
The dynamics of these curves are related to the main concern of the models applied in this article: each time that the curves of projected and observed rates are more separated, the model risk may be higher.
In the worst cases the distances between those curves may be systematic and increasing.

\begin{table}
{\small
\begin{center}
\begin{tabular}{llrrcrrcrrcrr}
\hline
 & & \multicolumn{5}{c}{Female} & & \multicolumn{5}{c}{Male} \\
\cline{3-7}\cline{9-13}
Country & Model & \multicolumn{2}{c}{Fitt. Period} & & \multicolumn{2}{c}{Frcs. Period} & & \multicolumn{2}{c}{Fitt. Period} & & \multicolumn{2}{c}{Frcs. Period} \\
\cline{3-4}\cline{6-7}\cline{9-10}\cline{12-13}
 & & \makebox[0.55cm][c]{\centering MSE$^*$} & \makebox[0.55cm][c]{\centering MAPE} & & \makebox[0.55cm][c]{\centering MSE$^*$} & \makebox[0.55cm][c]{\centering MAPE} & & \makebox[0.55cm][c]{\centering MSE$^*$} & \makebox[0.55cm][c]{\centering MAPE} & & \makebox[0.55cm][c]{\centering MSE$^*$} & \makebox[0.55cm][c]{\centering MAPE} \\
\hline
BE  & CBD  & \cellcolor[gray]{1.0} 0.07  & \cellcolor[gray]{1.0} 2.4  & & \cellcolor[gray]{0.8} 0.09  & \cellcolor[gray]{1.0} 5.3  & & \cellcolor[gray]{0.8} 0.17  & \cellcolor[gray]{0.8} 2.7  & & \cellcolor[gray]{0.8} 1.17  & \cellcolor[gray]{0.8} 12.1  \\
 & LC  & \cellcolor[gray]{0.8} 0.07  & \cellcolor[gray]{0.8} 2.3  & & \cellcolor[gray]{1.0} 0.10  & \cellcolor[gray]{0.8} 4.5  & & \cellcolor[gray]{1.0} 0.23  & \cellcolor[gray]{1.0} 2.8  & & \cellcolor[gray]{1.0} 1.2  & \cellcolor[gray]{1.0} 12.6  \\
 & SL  & \cellcolor[gray]{1.0} 0.07  & \cellcolor[gray]{1.0} 2.5  & & \cellcolor[gray]{1.0} 0.16  & \cellcolor[gray]{1.0} 6.1  & & \cellcolor[gray]{1.0} 0.59  & \cellcolor[gray]{1.0} 3.8  & & \cellcolor[gray]{1.0} 1.7  & \cellcolor[gray]{1.0} 14.2  \\
\hline
FR  & CBD  & \cellcolor[gray]{1.0} 0.03  & \cellcolor[gray]{1.0} 2.2  & & \cellcolor[gray]{0.8} 0.07  & \cellcolor[gray]{1.0} 6.8  & & \cellcolor[gray]{0.8} 0.06  & \cellcolor[gray]{1.0} 2.1  & & \cellcolor[gray]{0.8} 0.30  & \cellcolor[gray]{1.0} 7.2  \\
 & LC  & \cellcolor[gray]{0.8} 0.03  & \cellcolor[gray]{1.0} 1.7  & & \cellcolor[gray]{1.0} 0.13  & \cellcolor[gray]{1.0} 7.2  & & \cellcolor[gray]{1.0} 0.10  & \cellcolor[gray]{1.0} 2.0  & & \cellcolor[gray]{1.0} 0.4  & \cellcolor[gray]{0.8} 6.8  \\
 & SL  & \cellcolor[gray]{1.0} 0.03  & \cellcolor[gray]{0.8} 1.6  & & \cellcolor[gray]{1.0} 0.09  & \cellcolor[gray]{0.8} 5.3  & & \cellcolor[gray]{1.0} 0.09  & \cellcolor[gray]{0.8} 1.9  & & \cellcolor[gray]{1.0} 0.4  & \cellcolor[gray]{1.0} 7.9  \\
\hline
IT  & CBD  & \cellcolor[gray]{1.0} 0.06  & \cellcolor[gray]{1.0} 2.3  & & \cellcolor[gray]{0.8} 0.08  & \cellcolor[gray]{1.0} 5.3  & & \cellcolor[gray]{1.0} 0.07  & \cellcolor[gray]{1.0} 2.1  & & \cellcolor[gray]{0.8} 0.46  & \cellcolor[gray]{0.8} 10.3  \\
 & LC  & \cellcolor[gray]{0.8} 0.04  & \cellcolor[gray]{0.8} 2.0  & & \cellcolor[gray]{1.0} 0.13  & \cellcolor[gray]{0.8} 4.7  & & \cellcolor[gray]{1.0} 0.12  & \cellcolor[gray]{1.0} 2.3  & & \cellcolor[gray]{1.0} 0.6  & \cellcolor[gray]{1.0} 10.3  \\
 & SL  & \cellcolor[gray]{1.0} 0.05  & \cellcolor[gray]{1.0} 2.0  & & \cellcolor[gray]{1.0} 0.17  & \cellcolor[gray]{1.0} 6.9  & & \cellcolor[gray]{0.8} 0.07  & \cellcolor[gray]{0.8} 2.1  & & \cellcolor[gray]{1.0} 0.6  & \cellcolor[gray]{1.0} 11.1  \\
\hline
JP  & CBD  & \cellcolor[gray]{1.0} 0.07  & \cellcolor[gray]{1.0} 2.2  & & \cellcolor[gray]{1.0} 0.14  & \cellcolor[gray]{1.0} 9.1  & & \cellcolor[gray]{1.0} 0.10  & \cellcolor[gray]{1.0} 1.8  & & \cellcolor[gray]{1.0} 0.11  & \cellcolor[gray]{1.0} 8.7  \\
 & LC  & \cellcolor[gray]{0.8} 0.04  & \cellcolor[gray]{1.0} 1.8  & & \cellcolor[gray]{1.0} 0.32  & \cellcolor[gray]{1.0} 11.3  & & \cellcolor[gray]{0.8} 0.07  & \cellcolor[gray]{0.8} 1.5  & & \cellcolor[gray]{0.8} 0.1  & \cellcolor[gray]{1.0} 9.3  \\
 & SL  & \cellcolor[gray]{1.0} 0.04  & \cellcolor[gray]{0.8} 1.6  & & \cellcolor[gray]{0.8} 0.10  & \cellcolor[gray]{0.8} 7.9  & & \cellcolor[gray]{1.0} 0.07  & \cellcolor[gray]{1.0} 1.5  & & \cellcolor[gray]{1.0} 0.2  & \cellcolor[gray]{0.8} 8.7  \\
\hline
SW  & CBD  & \cellcolor[gray]{1.0} 0.10  & \cellcolor[gray]{1.0} 3.0  & & \cellcolor[gray]{1.0} 0.36  & \cellcolor[gray]{1.0} 8.5  & & \cellcolor[gray]{0.8} 0.09  & \cellcolor[gray]{0.8} 2.2  & & \cellcolor[gray]{0.8} 0.35  & \cellcolor[gray]{0.8} 9.1  \\
 & LC  & \cellcolor[gray]{0.8} 0.06  & \cellcolor[gray]{0.8} 2.5  & & \cellcolor[gray]{1.0} 0.22  & \cellcolor[gray]{1.0} 7.8  & & \cellcolor[gray]{1.0} 0.13  & \cellcolor[gray]{1.0} 2.3  & & \cellcolor[gray]{1.0} 0.5  & \cellcolor[gray]{1.0} 9.7  \\
 & SL  & \cellcolor[gray]{1.0} 0.12  & \cellcolor[gray]{1.0} 3.0  & & \cellcolor[gray]{0.8} 0.19  & \cellcolor[gray]{0.8} 7.8  & & \cellcolor[gray]{1.0} 0.14  & \cellcolor[gray]{1.0} 2.5  & & \cellcolor[gray]{1.0} 0.5  & \cellcolor[gray]{1.0} 10.3  \\
\hline
UK  & CBD  & \cellcolor[gray]{1.0} 0.03  & \cellcolor[gray]{1.0} 1.9  & & \cellcolor[gray]{1.0} 0.15  & \cellcolor[gray]{1.0} 8.3  & & \cellcolor[gray]{0.8} 0.06  & \cellcolor[gray]{0.8} 1.6  & & \cellcolor[gray]{1.0} 1.98  & \cellcolor[gray]{1.0} 14.8  \\
 & LC  & \cellcolor[gray]{0.8} 0.02  & \cellcolor[gray]{0.8} 1.4  & & \cellcolor[gray]{1.0} 0.13  & \cellcolor[gray]{1.0} 7.5  & & \cellcolor[gray]{1.0} 0.09  & \cellcolor[gray]{1.0} 1.7  & & \cellcolor[gray]{1.0} 1.7  & \cellcolor[gray]{1.0} 14.6  \\
 & SL  & \cellcolor[gray]{1.0} 0.03  & \cellcolor[gray]{1.0} 1.6  & & \cellcolor[gray]{0.8} 0.09  & \cellcolor[gray]{0.8} 6.8  & & \cellcolor[gray]{1.0} 0.11  & \cellcolor[gray]{1.0} 2.0  & & \cellcolor[gray]{0.8} 1.2  & \cellcolor[gray]{0.8} 13.1  \\
\hline
US  & CBD  & \cellcolor[gray]{1.0} 0.04  & \cellcolor[gray]{1.0} 2.8  & & \cellcolor[gray]{1.0} 0.57  & \cellcolor[gray]{1.0} 7.4  & & \cellcolor[gray]{1.0} 0.05  & \cellcolor[gray]{1.0} 1.9  & & \cellcolor[gray]{0.8} 0.31  & \cellcolor[gray]{0.8} 7.1  \\
 & LC  & \cellcolor[gray]{0.8} 0.01  & \cellcolor[gray]{0.8} 1.5  & & \cellcolor[gray]{1.0} 0.36  & \cellcolor[gray]{1.0} 7.1  & & \cellcolor[gray]{0.8} 0.05  & \cellcolor[gray]{0.8} 1.7  & & \cellcolor[gray]{1.0} 0.4  & \cellcolor[gray]{1.0} 7.6  \\
 & SL  & \cellcolor[gray]{1.0} 0.03  & \cellcolor[gray]{1.0} 1.9  & & \cellcolor[gray]{0.8} 0.30  & \cellcolor[gray]{0.8} 6.5  & & \cellcolor[gray]{1.0} 0.08  & \cellcolor[gray]{1.0} 2.0  & & \cellcolor[gray]{1.0} 0.5  & \cellcolor[gray]{1.0} 8.3  \\
\hline
\multicolumn{12}{l}{\quad MSE$^*$ = 10,000 $\times$ MSE.}
\end{tabular}
\end{center}
}
\caption{\label{tab01_65196011_1} MSE and MAPE of $q_{x,t}$ for $x_{\min}$ = 65 and  $x_{\max}$ = 89}
\end{table}


On females, Figure \ref{fig11}, the observed rates for IT and US present little variation, and are overlapped by predicted rates given by the CBD and LC models for IT and the LC and SL models for US.
On BE, the forecasted rates of the CBD and SL models tend to overlap the observed ones, whereas the forecasted rates of the LC model lightly overestimates the observed ones.
For FR, JP, SW, and UK, the observed rates are hard to represent by the benchmark and new models since systematic differences between observed and predicted rates are found.

The behaviors of the forecasted rates for males, Figure \ref{fig12}, are quite different from those for females, Figure \ref{fig11}.
On males, we now have through all the countries studied in this paper (excepting JP) that the forecasted rates tend to underestimate the observed rates.
The best forecasted rates are found for FR and US, with the LC model for FR and the LS model for US.
For the other countries the forecasted rates are away from the observed rates, these differences growing when time increases.

\begin{table}
{\small
\begin{center}
\begin{tabular}{llrrcrrcrrcrr}
\hline
 & & \multicolumn{5}{c}{Female} & & \multicolumn{5}{c}{Male} \\
\cline{3-7}\cline{9-13}
Country & Model & \multicolumn{2}{c}{Fitt. Period} & & \multicolumn{2}{c}{Frcs. Period} & & \multicolumn{2}{c}{Fitt. Period} & & \multicolumn{2}{c}{Frcs. Period} \\
\cline{3-4}\cline{6-7}\cline{9-10}\cline{12-13}
 & & \makebox[0.55cm][c]{\centering MSE$^*$} & \makebox[0.55cm][c]{\centering MAPE} & & \makebox[0.55cm][c]{\centering MSE$^*$} & \makebox[0.55cm][c]{\centering MAPE} & & \makebox[0.55cm][c]{\centering MSE$^*$} & \makebox[0.55cm][c]{\centering MAPE} & & \makebox[0.55cm][c]{\centering MSE$^*$} & \makebox[0.55cm][c]{\centering MAPE} \\
\hline
BE  & CBD  & \cellcolor[gray]{1.0} 0.34  & \cellcolor[gray]{1.0} 3.6  & & \cellcolor[gray]{0.8} 0.20  & \cellcolor[gray]{1.0} 7.5  & & \cellcolor[gray]{0.8} 0.48  & \cellcolor[gray]{1.0} 3.3  & & \cellcolor[gray]{0.8} 1.03  & \cellcolor[gray]{0.8} 10.0  \\
 & LC  & \cellcolor[gray]{0.8} 0.23  & \cellcolor[gray]{0.8} 2.6  & & \cellcolor[gray]{1.0} 0.26  & \cellcolor[gray]{0.8} 5.0  & & \cellcolor[gray]{1.0} 0.55  & \cellcolor[gray]{0.8} 3.2  & & \cellcolor[gray]{1.0} 1.9  & \cellcolor[gray]{1.0} 11.8  \\
 & SL  & \cellcolor[gray]{1.0} 0.25  & \cellcolor[gray]{1.0} 2.9  & & \cellcolor[gray]{1.0} 0.41  & \cellcolor[gray]{1.0} 6.9  & & \cellcolor[gray]{1.0} 0.98  & \cellcolor[gray]{1.0} 3.9  & & \cellcolor[gray]{1.0} 2.6  & \cellcolor[gray]{1.0} 14.0  \\
\hline
FR  & CBD  & \cellcolor[gray]{1.0} 0.15  & \cellcolor[gray]{1.0} 4.1  & & \cellcolor[gray]{0.8} 0.19  & \cellcolor[gray]{1.0} 9.8  & & \cellcolor[gray]{0.8} 0.18  & \cellcolor[gray]{1.0} 3.1  & & \cellcolor[gray]{1.0} 0.69  & \cellcolor[gray]{1.0} 8.7  \\
 & LC  & \cellcolor[gray]{1.0} 0.06  & \cellcolor[gray]{1.0} 1.9  & & \cellcolor[gray]{1.0} 0.20  & \cellcolor[gray]{1.0} 7.1  & & \cellcolor[gray]{1.0} 0.21  & \cellcolor[gray]{0.8} 2.3  & & \cellcolor[gray]{0.8} 0.5  & \cellcolor[gray]{0.8} 6.6  \\
 & SL  & \cellcolor[gray]{0.8} 0.05  & \cellcolor[gray]{0.8} 1.8  & & \cellcolor[gray]{1.0} 0.21  & \cellcolor[gray]{0.8} 6.1  & & \cellcolor[gray]{1.0} 0.40  & \cellcolor[gray]{1.0} 2.5  & & \cellcolor[gray]{1.0} 1.3  & \cellcolor[gray]{1.0} 9.1  \\
\hline
IT  & CBD  & \cellcolor[gray]{1.0} 0.26  & \cellcolor[gray]{1.0} 3.5  & & \cellcolor[gray]{1.0} 0.19  & \cellcolor[gray]{1.0} 6.5  & & \cellcolor[gray]{0.8} 0.17  & \cellcolor[gray]{1.0} 2.6  & & \cellcolor[gray]{0.8} 0.61  & \cellcolor[gray]{0.8} 11.1  \\
 & LC  & \cellcolor[gray]{1.0} 0.11  & \cellcolor[gray]{1.0} 2.1  & & \cellcolor[gray]{0.8} 0.15  & \cellcolor[gray]{0.8} 4.3  & & \cellcolor[gray]{1.0} 0.25  & \cellcolor[gray]{1.0} 2.5  & & \cellcolor[gray]{1.0} 0.8  & \cellcolor[gray]{1.0} 11.2  \\
 & SL  & \cellcolor[gray]{0.8} 0.09  & \cellcolor[gray]{0.8} 2.1  & & \cellcolor[gray]{1.0} 0.39  & \cellcolor[gray]{1.0} 6.0  & & \cellcolor[gray]{1.0} 0.21  & \cellcolor[gray]{0.8} 2.4  & & \cellcolor[gray]{1.0} 0.8  & \cellcolor[gray]{1.0} 11.7  \\
\hline
JP  & CBD  & \cellcolor[gray]{1.0} 0.28  & \cellcolor[gray]{1.0} 3.5  & & \cellcolor[gray]{0.8} 0.23  & \cellcolor[gray]{1.0} 11.1  & & \cellcolor[gray]{1.0} 0.45  & \cellcolor[gray]{1.0} 3.0  & & \cellcolor[gray]{1.0} 0.23  & \cellcolor[gray]{0.8} 8.9  \\
 & LC  & \cellcolor[gray]{0.8} 0.11  & \cellcolor[gray]{1.0} 2.0  & & \cellcolor[gray]{1.0} 0.68  & \cellcolor[gray]{1.0} 13.4  & & \cellcolor[gray]{1.0} 0.20  & \cellcolor[gray]{0.8} 1.9  & & \cellcolor[gray]{0.8} 0.2  & \cellcolor[gray]{1.0} 10.2  \\
 & SL  & \cellcolor[gray]{1.0} 0.17  & \cellcolor[gray]{0.8} 2.0  & & \cellcolor[gray]{1.0} 0.23  & \cellcolor[gray]{0.8} 9.3  & & \cellcolor[gray]{0.8} 0.20  & \cellcolor[gray]{1.0} 2.0  & & \cellcolor[gray]{1.0} 1.1  & \cellcolor[gray]{1.0} 9.1  \\
\hline
SW  & CBD  & \cellcolor[gray]{1.0} 0.43  & \cellcolor[gray]{1.0} 4.3  & & \cellcolor[gray]{1.0} 1.20  & \cellcolor[gray]{1.0} 10.4  & & \cellcolor[gray]{0.8} 0.39  & \cellcolor[gray]{0.8} 2.7  & & \cellcolor[gray]{0.8} 0.61  & \cellcolor[gray]{0.8} 9.2  \\
 & LC  & \cellcolor[gray]{0.8} 0.20  & \cellcolor[gray]{0.8} 2.9  & & \cellcolor[gray]{0.8} 0.71  & \cellcolor[gray]{0.8} 7.9  & & \cellcolor[gray]{1.0} 0.50  & \cellcolor[gray]{1.0} 2.8  & & \cellcolor[gray]{1.0} 1.6  & \cellcolor[gray]{1.0} 10.8  \\
 & SL  & \cellcolor[gray]{1.0} 0.29  & \cellcolor[gray]{1.0} 3.3  & & \cellcolor[gray]{1.0} 1.15  & \cellcolor[gray]{1.0} 9.0  & & \cellcolor[gray]{1.0} 0.72  & \cellcolor[gray]{1.0} 3.2  & & \cellcolor[gray]{1.0} 2.0  & \cellcolor[gray]{1.0} 11.2  \\
\hline
UK  & CBD  & \cellcolor[gray]{1.0} 0.11  & \cellcolor[gray]{1.0} 2.7  & & \cellcolor[gray]{1.0} 0.30  & \cellcolor[gray]{1.0} 9.7  & & \cellcolor[gray]{0.8} 0.14  & \cellcolor[gray]{1.0} 2.3  & & \cellcolor[gray]{1.0} 1.84  & \cellcolor[gray]{1.0} 13.8  \\
 & LC  & \cellcolor[gray]{0.8} 0.05  & \cellcolor[gray]{0.8} 1.6  & & \cellcolor[gray]{1.0} 0.16  & \cellcolor[gray]{1.0} 8.8  & & \cellcolor[gray]{1.0} 0.16  & \cellcolor[gray]{0.8} 2.0  & & \cellcolor[gray]{1.0} 1.3  & \cellcolor[gray]{1.0} 13.6  \\
 & SL  & \cellcolor[gray]{1.0} 0.09  & \cellcolor[gray]{1.0} 1.9  & & \cellcolor[gray]{0.8} 0.15  & \cellcolor[gray]{0.8} 7.8  & & \cellcolor[gray]{1.0} 0.19  & \cellcolor[gray]{1.0} 2.1  & & \cellcolor[gray]{0.8} 1.1  & \cellcolor[gray]{0.8} 12.5  \\
\hline
US  & CBD  & \cellcolor[gray]{1.0} 0.11  & \cellcolor[gray]{1.0} 4.1  & & \cellcolor[gray]{1.0} 1.73  & \cellcolor[gray]{1.0} 8.1  & & \cellcolor[gray]{0.8} 0.07  & \cellcolor[gray]{1.0} 2.0  & & \cellcolor[gray]{0.8} 0.57  & \cellcolor[gray]{0.8} 7.2  \\
 & LC  & \cellcolor[gray]{0.8} 0.02  & \cellcolor[gray]{0.8} 1.6  & & \cellcolor[gray]{0.8} 0.93  & \cellcolor[gray]{1.0} 7.4  & & \cellcolor[gray]{1.0} 0.08  & \cellcolor[gray]{0.8} 1.7  & & \cellcolor[gray]{1.0} 0.8  & \cellcolor[gray]{1.0} 7.8  \\
 & SL  & \cellcolor[gray]{1.0} 0.05  & \cellcolor[gray]{1.0} 2.0  & & \cellcolor[gray]{1.0} 0.95  & \cellcolor[gray]{0.8} 6.9  & & \cellcolor[gray]{1.0} 0.09  & \cellcolor[gray]{1.0} 2.0  & & \cellcolor[gray]{1.0} 0.8  & \cellcolor[gray]{1.0} 8.0  \\
\hline
\multicolumn{12}{l}{\quad MSE$^*$ = 10,000 $\times$ MSE.}
\end{tabular}
\end{center}
}
\caption{\label{tab01_60196011_2} MSE and MAPE of $q_{x,t}$ for $x_{\min}$ = 60 and  $x_{\max}$ = 94}
\end{table}


\begin{table}
{\small
\begin{center}
\begin{tabular}{llrrcrrcrrcrr}
\hline
 & & \multicolumn{5}{c}{Female} & & \multicolumn{5}{c}{Male} \\
\cline{3-7}\cline{9-13}
Country & Model & \multicolumn{2}{c}{Fitt. Period} & & \multicolumn{2}{c}{Frcs. Period} & & \multicolumn{2}{c}{Fitt. Period} & & \multicolumn{2}{c}{Frcs. Period} \\
\cline{3-4}\cline{6-7}\cline{9-10}\cline{12-13}
 & & \makebox[0.55cm][c]{\centering MSE$^*$} & \makebox[0.55cm][c]{\centering MAPE} & & \makebox[0.55cm][c]{\centering MSE$^*$} & \makebox[0.55cm][c]{\centering MAPE} & & \makebox[0.55cm][c]{\centering MSE$^*$} & \makebox[0.55cm][c]{\centering MAPE} & & \makebox[0.55cm][c]{\centering MSE$^*$} & \makebox[0.55cm][c]{\centering MAPE} \\
\hline
BE  & CBD  & \cellcolor[gray]{1.0} 0.41  & \cellcolor[gray]{1.0} 3.2  & & \cellcolor[gray]{0.8} 0.21  & \cellcolor[gray]{1.0} 5.0  & & \cellcolor[gray]{0.8} 0.52  & \cellcolor[gray]{0.8} 3.0  & & \cellcolor[gray]{0.8} 1.17  & \cellcolor[gray]{0.8} 10.5  \\
 & LC  & \cellcolor[gray]{0.8} 0.27  & \cellcolor[gray]{0.8} 2.5  & & \cellcolor[gray]{1.0} 0.31  & \cellcolor[gray]{0.8} 4.8  & & \cellcolor[gray]{1.0} 0.64  & \cellcolor[gray]{1.0} 3.0  & & \cellcolor[gray]{1.0} 2.2  & \cellcolor[gray]{1.0} 12.3  \\
 & SL  & \cellcolor[gray]{1.0} 0.30  & \cellcolor[gray]{1.0} 2.7  & & \cellcolor[gray]{1.0} 0.45  & \cellcolor[gray]{1.0} 6.6  & & \cellcolor[gray]{1.0} 1.03  & \cellcolor[gray]{1.0} 4.0  & & \cellcolor[gray]{1.0} 3.2  & \cellcolor[gray]{1.0} 13.4  \\
\hline
FR  & CBD  & \cellcolor[gray]{1.0} 0.21  & \cellcolor[gray]{1.0} 2.8  & & \cellcolor[gray]{0.8} 0.16  & \cellcolor[gray]{1.0} 6.5  & & \cellcolor[gray]{0.8} 0.16  & \cellcolor[gray]{0.8} 2.1  & & \cellcolor[gray]{0.8} 0.45  & \cellcolor[gray]{1.0} 6.8  \\
 & LC  & \cellcolor[gray]{1.0} 0.07  & \cellcolor[gray]{1.0} 1.7  & & \cellcolor[gray]{1.0} 0.22  & \cellcolor[gray]{1.0} 6.5  & & \cellcolor[gray]{1.0} 0.24  & \cellcolor[gray]{1.0} 2.1  & & \cellcolor[gray]{1.0} 0.5  & \cellcolor[gray]{0.8} 6.2  \\
 & SL  & \cellcolor[gray]{0.8} 0.06  & \cellcolor[gray]{0.8} 1.6  & & \cellcolor[gray]{1.0} 0.22  & \cellcolor[gray]{0.8} 5.9  & & \cellcolor[gray]{1.0} 0.46  & \cellcolor[gray]{1.0} 2.3  & & \cellcolor[gray]{1.0} 1.7  & \cellcolor[gray]{1.0} 8.9  \\
\hline
IT  & CBD  & \cellcolor[gray]{1.0} 0.33  & \cellcolor[gray]{1.0} 3.2  & & \cellcolor[gray]{0.8} 0.14  & \cellcolor[gray]{1.0} 5.0  & & \cellcolor[gray]{0.8} 0.18  & \cellcolor[gray]{1.0} 2.3  & & \cellcolor[gray]{0.8} 0.60  & \cellcolor[gray]{0.8} 9.3  \\
 & LC  & \cellcolor[gray]{1.0} 0.12  & \cellcolor[gray]{1.0} 2.0  & & \cellcolor[gray]{1.0} 0.17  & \cellcolor[gray]{0.8} 4.3  & & \cellcolor[gray]{1.0} 0.29  & \cellcolor[gray]{1.0} 2.4  & & \cellcolor[gray]{1.0} 1.0  & \cellcolor[gray]{1.0} 9.8  \\
 & SL  & \cellcolor[gray]{0.8} 0.12  & \cellcolor[gray]{0.8} 2.0  & & \cellcolor[gray]{1.0} 0.54  & \cellcolor[gray]{1.0} 6.9  & & \cellcolor[gray]{1.0} 0.21  & \cellcolor[gray]{0.8} 2.2  & & \cellcolor[gray]{1.0} 0.8  & \cellcolor[gray]{1.0} 10.0  \\
\hline
JP  & CBD  & \cellcolor[gray]{1.0} 0.34  & \cellcolor[gray]{1.0} 3.1  & & \cellcolor[gray]{1.0} 0.51  & \cellcolor[gray]{1.0} 8.6  & & \cellcolor[gray]{1.0} 0.46  & \cellcolor[gray]{1.0} 2.8  & & \cellcolor[gray]{1.0} 0.22  & \cellcolor[gray]{1.0} 8.0  \\
 & LC  & \cellcolor[gray]{0.8} 0.12  & \cellcolor[gray]{1.0} 1.9  & & \cellcolor[gray]{1.0} 0.75  & \cellcolor[gray]{1.0} 11.0  & & \cellcolor[gray]{1.0} 0.22  & \cellcolor[gray]{1.0} 1.7  & & \cellcolor[gray]{0.8} 0.2  & \cellcolor[gray]{1.0} 8.3  \\
 & SL  & \cellcolor[gray]{1.0} 0.16  & \cellcolor[gray]{0.8} 1.7  & & \cellcolor[gray]{0.8} 0.27  & \cellcolor[gray]{0.8} 8.1  & & \cellcolor[gray]{0.8} 0.20  & \cellcolor[gray]{0.8} 1.7  & & \cellcolor[gray]{1.0} 0.6  & \cellcolor[gray]{0.8} 7.9  \\
\hline
SW  & CBD  & \cellcolor[gray]{1.0} 0.48  & \cellcolor[gray]{1.0} 3.8  & & \cellcolor[gray]{1.0} 0.95  & \cellcolor[gray]{1.0} 8.6  & & \cellcolor[gray]{0.8} 0.43  & \cellcolor[gray]{0.8} 2.6  & & \cellcolor[gray]{0.8} 0.70  & \cellcolor[gray]{0.8} 8.5  \\
 & LC  & \cellcolor[gray]{0.8} 0.23  & \cellcolor[gray]{0.8} 2.7  & & \cellcolor[gray]{1.0} 0.80  & \cellcolor[gray]{1.0} 7.8  & & \cellcolor[gray]{1.0} 0.58  & \cellcolor[gray]{1.0} 2.7  & & \cellcolor[gray]{1.0} 2.0  & \cellcolor[gray]{1.0} 10.3  \\
 & SL  & \cellcolor[gray]{1.0} 0.27  & \cellcolor[gray]{1.0} 3.0  & & \cellcolor[gray]{0.8} 0.63  & \cellcolor[gray]{0.8} 7.6  & & \cellcolor[gray]{1.0} 0.81  & \cellcolor[gray]{1.0} 3.1  & & \cellcolor[gray]{1.0} 1.9  & \cellcolor[gray]{1.0} 10.5  \\
\hline
UK  & CBD  & \cellcolor[gray]{1.0} 0.12  & \cellcolor[gray]{1.0} 2.2  & & \cellcolor[gray]{1.0} 0.25  & \cellcolor[gray]{1.0} 7.7  & & \cellcolor[gray]{0.8} 0.13  & \cellcolor[gray]{1.0} 1.8  & & \cellcolor[gray]{1.0} 2.03  & \cellcolor[gray]{1.0} 13.2  \\
 & LC  & \cellcolor[gray]{0.8} 0.06  & \cellcolor[gray]{0.8} 1.5  & & \cellcolor[gray]{1.0} 0.18  & \cellcolor[gray]{1.0} 6.8  & & \cellcolor[gray]{1.0} 0.18  & \cellcolor[gray]{0.8} 1.8  & & \cellcolor[gray]{1.0} 1.6  & \cellcolor[gray]{1.0} 12.8  \\
 & SL  & \cellcolor[gray]{1.0} 0.13  & \cellcolor[gray]{1.0} 1.8  & & \cellcolor[gray]{0.8} 0.15  & \cellcolor[gray]{0.8} 6.1  & & \cellcolor[gray]{1.0} 0.19  & \cellcolor[gray]{1.0} 2.0  & & \cellcolor[gray]{0.8} 1.3  & \cellcolor[gray]{0.8} 11.8  \\
\hline
US  & CBD  & \cellcolor[gray]{1.0} 0.07  & \cellcolor[gray]{1.0} 2.8  & & \cellcolor[gray]{1.0} 1.61  & \cellcolor[gray]{1.0} 8.3  & & \cellcolor[gray]{0.8} 0.07  & \cellcolor[gray]{1.0} 1.8  & & \cellcolor[gray]{0.8} 0.59  & \cellcolor[gray]{0.8} 6.8  \\
 & LC  & \cellcolor[gray]{0.8} 0.02  & \cellcolor[gray]{0.8} 1.4  & & \cellcolor[gray]{1.0} 1.07  & \cellcolor[gray]{1.0} 7.9  & & \cellcolor[gray]{1.0} 0.09  & \cellcolor[gray]{0.8} 1.7  & & \cellcolor[gray]{1.0} 1.0  & \cellcolor[gray]{1.0} 8.0  \\
 & SL  & \cellcolor[gray]{1.0} 0.09  & \cellcolor[gray]{1.0} 2.0  & & \cellcolor[gray]{0.8} 0.98  & \cellcolor[gray]{0.8} 7.5  & & \cellcolor[gray]{1.0} 0.12  & \cellcolor[gray]{1.0} 2.0  & & \cellcolor[gray]{1.0} 1.1  & \cellcolor[gray]{1.0} 8.5  \\
\hline
\multicolumn{12}{l}{\quad MSE$^*$ = 10,000 $\times$ MSE.}
\end{tabular}
\end{center}
}
\caption{\label{tab01_65196011_2} MSE and MAPE of $q_{x,t}$ for $x_{\min}$ = 65 and  $x_{\max}$ = 94}
\end{table}

The projected and observed curves shown in Figures \ref{fig11} and \ref{fig12} are in general more separated in the last years of the forecasting period.
Let us examine the differences among these curves in the two last years, 2008 and 2009, computing the next index of type MAPE.
$$
\textrm{MAPE}_{\Delta_{65,2008,2009}}=\frac{1}{2}\left(\frac{\big|\Delta_{65,2008}-\hat{\Delta}_{65,2008}\big|}{\hat{\Delta}_{65,2008}}
+\frac{\big|\Delta_{65,2009}-\hat{\Delta}_{65,2009}\big|}{\hat{\Delta}_{65,2009}}\right)\times100~\%\textrm{.}
$$
%
Results of this index are shown in Table \ref{tab02}.
They are organized by country, sex, and $x_{\min}$ and $x_{\max}$, so they contain the cases considered in Figures \ref{fig11} and \ref{fig12}.
The lowest values of $\textrm{MAPE}_{\Delta_{65,2008,2009}}$ among the models studied are highlighted.
Smaller highlighted values indicate a better forecast of the data of 2008 and 2009, and a few of them are found only, for instance, considering those near 10.0~\% or less, we have IT and US females and FR and US males, all of them representing 21.4~\% of the cases.
%
Besides, the SL model seems to be appropriate for mortality forecasting for US females aged 65 and over and for US males aged 60 and over.
For US females aged 60 and over the forecasts given by the LS model seem also acceptable.

\begin{table}
{\small
\begin{center}
\begin{tabular}{lccrrrcrrr}
\hline
Country & $x_{\min}$ & $x_{\max}$ & \multicolumn{3}{c}{Female} & & \multicolumn{3}{c}{Male} \\
\cline{4-6}\cline{8-10}
 & & & \multicolumn{1}{c}{CBD} & \multicolumn{1}{c}{LC} & \multicolumn{1}{c}{SL} & & \multicolumn{1}{c}{CBD} & \multicolumn{1}{c}{LC} & \multicolumn{1}{c}{SL} \\
\hline
BE  & 60  & 89  &  50.7  &  61.5  & \cellcolor[gray]{0.8} 40.8  & &  40.2  & \cellcolor[gray]{0.8} 34.9  &  51.4  \\
 & & 94  &  54.6  &  57.6  & \cellcolor[gray]{0.8} 40.6  & &  42.5  & \cellcolor[gray]{0.8} 36.8  &  52.9  \\
\cline{2-10}
 & 65  & 89  &  87.3  &  61.2  & \cellcolor[gray]{0.8} 43.1  & &  47.8  & \cellcolor[gray]{0.8} 39.2  &  52.7  \\
 & & 94  &  89.0  &  57.0  & \cellcolor[gray]{0.8} 42.8  & &  51.7  & \cellcolor[gray]{0.8} 41.1  &  63.7  \\
\hline
FR  & 60  & 89  &  42.8  &  52.5  & \cellcolor[gray]{0.8} 27.9  & &  31.6  & \cellcolor[gray]{0.8} 19.5  &  35.6  \\
 & & 94  &  49.1  &  51.2  & \cellcolor[gray]{0.8} 28.2  & &  28.2  & \cellcolor[gray]{0.8} 19.4  &  33.2  \\
\cline{2-10}
 & 65  & 89  &  92.6  &  53.8  & \cellcolor[gray]{0.8} 31.4  & & \cellcolor[gray]{0.8} 5.7  &  17.0  &  35.1  \\
 & & 94  &  97.2  &  52.2  & \cellcolor[gray]{0.8} 30.5  & & \cellcolor[gray]{0.8} 3.1  &  17.0  &  33.8  \\
\hline
IT  & 60  & 89  &  8.6  & \cellcolor[gray]{0.8} 5.7  &  15.7  & &  63.6  & \cellcolor[gray]{0.8} 53.4  &  67.8  \\
 & & 94  &  6.3  & \cellcolor[gray]{0.8} 5.4  &  17.3  & &  64.1  & \cellcolor[gray]{0.8} 53.4  &  70.5  \\
\cline{2-10}
 & 65  & 89  &  7.3  & \cellcolor[gray]{0.8} 5.7  &  17.1  & &  59.3  & \cellcolor[gray]{0.8} 54.0  &  73.8  \\
 & & 94  &  9.5  & \cellcolor[gray]{0.8} 5.5  &  19.2  & &  61.1  & \cellcolor[gray]{0.8} 53.7  &  74.2  \\
\hline
JP  & 60  & 89  &  58.0  &  65.3  & \cellcolor[gray]{0.8} 47.6  & &  85.5  &  114.9  & \cellcolor[gray]{0.8} 72.5  \\
 & & 94  &  59.8  &  66.4  & \cellcolor[gray]{0.8} 48.6  & &  84.9  &  115.6  & \cellcolor[gray]{0.8} 78.3  \\
\cline{2-10}
 & 65  & 89  &  89.5  &  66.0  & \cellcolor[gray]{0.8} 56.0  & &  125.7  &  120.5  & \cellcolor[gray]{0.8} 104.1  \\
 & & 94  &  87.5  &  66.9  & \cellcolor[gray]{0.8} 57.0  & &  116.9  &  120.8  & \cellcolor[gray]{0.8} 107.1  \\
\hline
SW  & 60  & 89  &  70.7  & \cellcolor[gray]{0.8} 30.4  &  50.7  & & \cellcolor[gray]{0.8} 57.9  &  68.8  &  69.8  \\
 & & 94  &  76.9  & \cellcolor[gray]{0.8} 28.8  &  44.0  & & \cellcolor[gray]{0.8} 57.7  &  71.0  &  71.5  \\
\cline{2-10}
 & 65  & 89  &  109.3  &  29.5  & \cellcolor[gray]{0.8} 25.1  & & \cellcolor[gray]{0.8} 58.6  &  69.5  &  74.6  \\
 & & 94  &  115.9  & \cellcolor[gray]{0.8} 27.9  &  29.1  & & \cellcolor[gray]{0.8} 58.2  &  72.1  &  74.1  \\
\hline
UK  & 60  & 89  &  61.8  &  70.1  & \cellcolor[gray]{0.8} 60.7  & & \cellcolor[gray]{0.8} 62.5  &  63.1  &  62.8  \\
 & & 94  & \cellcolor[gray]{0.8} 57.3  &  70.6  &  60.9  & &  63.7  &  63.8  & \cellcolor[gray]{0.8} 62.5  \\
\cline{2-10}
 & 65  & 89  & \cellcolor[gray]{0.8} 52.0  &  70.5  &  71.8  & &  66.5  &  64.9  & \cellcolor[gray]{0.8} 61.7  \\
 & & 94  & \cellcolor[gray]{0.8} 44.0  &  71.0  &  71.8  & &  68.4  &  65.8  & \cellcolor[gray]{0.8} 61.9  \\
\hline
US  & 60  & 89  &  9.6  & \cellcolor[gray]{0.8} 9.4  &  9.7  & &  33.5  &  22.7  & \cellcolor[gray]{0.8} 10.6  \\
 & & 94  &  14.4  & \cellcolor[gray]{0.8} 9.3  &  9.6  & &  33.0  &  23.6  & \cellcolor[gray]{0.8} 10.0  \\
\cline{2-10}
 & 65  & 89  &  28.6  &  9.3  & \cellcolor[gray]{0.8} 8.7  & &  28.7  &  21.8  & \cellcolor[gray]{0.8} 16.1  \\
 & & 94  &  45.3  &  9.2  & \cellcolor[gray]{0.8} 8.9  & &  28.5  &  23.0  & \cellcolor[gray]{0.8} 13.2  \\
\hline
\end{tabular}
\end{center}
}
\caption{\label{tab02} $\textrm{MAPE}_{\Delta_{65,2008,2009}}$}
\end{table}

\section{Conclusion}

A new model for stochastic mortality projection based on the application of the transform $\log(-\log x)$ to survival functions was proposed.
This transformation was represented by speci\-fic-age parameters and stochastic processes depending on time.
This model has a structure like those given in \cite{Brass1974} and \cite{deJongMarshall2007}.
Mortality forecasting was obtained from the projection of the time-processes.

According to some goodness-of-fit criteria the application of this model to sex-based mortality in-sample and out-of-sample data from seven countries showed that in some cases it overperforms two well-known stochastic mortality models.
These findings were sharp for females over the studied forecasting year period.
These global assessments where complemented with valuations per year of the rates of MI.
These last results gave appraisals of the mortality forecasting quality, showing that in a few cases forecasts of the rates of MI may be acceptable.
In many cases these mortality projections were away from the observed mortality.
These last findings corroborated the challenge that future mortality presents and showed the ``mortality gap'' between predicted and observed mortality that the new modelizations should mitigate.
For this aim, our new model seems to provide fundamental and enduring features of mortality patterns to deal with this mortality gap in some cases, these features being based on a reference year and on its relationships with subsequent years.
Hence, this new model seems a promising approach to give more accurate projections of mortality.

In many cases, the observed mortality gaps seem to tend to grow over years.
These model failures have been evidenced in the literature and they may produce longevity risk that may be refected in financial losses (see e.g. \cite{RichardsCurrie2009}).
Hence, the information provided by the new model may benefit practitioners in their efforts to reduce longevity risk in pricing and valuation of products involving longevity. 
%
A wide survey on impacts of longevity risk in pension funds and annuity providers is found in \cite{OECD}.

Another interesting result of our analysis is that the correspondence among the best models over both fitting and forecasting periods, i.e. that a same model be the best one over fitting and forecasting periods, was not systematic.
This finding expresses that the selection of models cannot reliably be based on the analysis of in-sample errors, as claimed by authors like e.g. Booth and Tickle in \cite{BoothTickle2008}, although evidently a model should provide a good fit to the historical data.
Further studies to analyze out-of-sample errors using historical data are required.

The results obtained in this paper depend on data features, namely length of the fitting year period, $t_0$, $x_{\min}$, and $x_{\max}$.
For instance, literature shows that shorter fitting periods would tend to work better because they capture the most recent mortality trend (see e.g. \cite{BoothHyndmanTickledeJong2006}).
Besides, according to our results, the selection of models would be impacted by variations of $x_{\min}$, $x_{\max}$, sex, and country.
This means that all these variables should be considered as parameters in mortality studies.

\section*{Acknowledgments} 
Meitner Cadena acknowledges the support of SWISS LIFE through its ESSEC research program on 'Consequences of the population ageing on the insurances loss'.




\end{document}